\documentclass[journal]{IEEEtran}
\IEEEoverridecommandlockouts

\pdfoutput=1

\usepackage[letterpaper, bottom=1in, right=0.625in, left=0.625in, top=0.75in]{geometry}
\usepackage{cite}
\usepackage{amsmath,amssymb,amsfonts,amsthm,bm}
\usepackage{graphicx}
\usepackage{textcomp}
\usepackage{xcolor}
\usepackage{caption}
\usepackage{subcaption}
\usepackage{enumitem}
\usepackage{cuted}
\usepackage{units}
\usepackage{multirow}
\usepackage[normalem]{ulem}
\usepackage[font=footnotesize,skip=1pt]{caption}
\usepackage[protrusion=true,expansion=true]{microtype}
\usepackage{balance}
\usepackage{mathtools}
\usepackage[linesnumbered,ruled,vlined]{algorithm2e}
\usepackage{algpseudocode}
\usepackage{array}
\usepackage{hhline}

\newtheorem{proposition}{Proposition}

\newtheorem{example}{Example}

\newtheorem{assumption}{Assumption}

\renewcommand{\qedsymbol}{$\blacksquare$}

\newcolumntype{C}{>{\centering\arraybackslash}p{0.625cm}}
\allowdisplaybreaks

\DeclareMathOperator*{\argmax}{arg\,max}
\def\BibTeX{{\rm B\kern-.05em{\sc i\kern-.025em b}\kern-.08em
    T\kern-.1667em\lower.7ex\hbox{E}\kern-.125emX}}

\begin{document}

\title{Minimum Description Feature Selection for Complexity Reduction in Machine Learning-based Wireless Positioning}

\author{Myeung~Suk~Oh,~\IEEEmembership{Graduate~Student~Member,~IEEE}, Anindya Bijoy Das,~\IEEEmembership{Member,~IEEE}, Taejoon~Kim,~\IEEEmembership{Senior~Member,~IEEE}, David~J.~Love,~\IEEEmembership{Fellow,~IEEE}, and Christopher~G.~Brinton,~\IEEEmembership{Senior~Member,~IEEE}
\thanks{An abridged version of this paper has been presented in the 2024 IEEE International Conference on Communications (ICC).}
\thanks{This work was supported in part by the National Science Foundation (NSF) under Grant CNS2146171, Grant CNS2212565, Grant CNS2225577, Grant CNS2225578, and Grant EEC1941529; and in part by the Office of Naval Research (ONR) under Grant N000142112472. \textit{(Corresponding author: Myeung Suk Oh.)}}
\thanks{Myeung Suk Oh, Anindya Bijoy Das, David J. Love, and Christopher G. Brinton are with the Elmore Family School of Electrical and Computer Engineering, Purdue University, West Lafayette, IN 47907 USA (e-mail: oh223@purdue.edu; das207@purdue.edu; djlove@purdue.edu; cgb@purdue.edu).}
\thanks{Taejoon Kim is with the School of Electrical, Computer and Energy Engineering, Arizona State University, Tempe, AZ  85287 USA (e-mail: taejoonkim@asu.edu).}
}

\maketitle 

\thispagestyle{empty}

\begin{abstract}
    Recently, deep learning approaches have provided solutions to difficult problems in wireless positioning (WP). Although these WP algorithms have attained excellent and consistent performance against complex channel environments, the computational complexity coming from processing high-dimensional features can be prohibitive for mobile applications. In this work, we design a novel positioning neural network (P-NN) that utilizes the minimum description features to substantially reduce the complexity of deep learning-based WP. P-NN's feature selection strategy is based on maximum power measurements and their temporal locations to convey information needed to conduct WP. We improve P-NN's learning ability by intelligently processing two different types of inputs: sparse image and measurement matrices. Specifically, we implement a self-attention layer to reinforce the training ability of our network. We also develop a technique to adapt feature space size, optimizing over the expected information gain and the classification capability quantified with information-theoretic measures on signal bin selection. Numerical results show that P-NN achieves a significant advantage in performance-complexity tradeoff over deep learning baselines that leverage the full power delay profile (PDP). In particular, we find that P-NN achieves a large improvement in performance for low SNR, as unnecessary measurements are discarded in our minimum description features.
\end{abstract}


\section{Introduction and Related Work}\label{sec:introduction}

An abundance of the today's mobile systems rely on the ability of devices to perceive and locate their surroundings.
Popular examples include object localization in autonomous vehicles~\cite{Kuutti20}, robotics~\cite{Wang19}, and unmanned aerial vehicles (UAVs)~\cite{Choi19}, as well as many other Internet of Things (IoT) use-cases~\cite{Arzo21}. 
Given the prevalence of wireless sensors in these systems, wireless positioning (WP) has become a commonly investigated technique for providing situational awareness in mobile applications.

WP is typically conducted using a group of wireless sensors that exchange signals with a target of interest in order to collect measurements that are informative for location estimation.
These sensors form a network, and the measurements from each sensor are collected by a data fusion center (DFC) for centralized processing to estimate the target location.
Among the types of signals that are popularly used for WP (e.g.,  Bluetooth~\cite{Cominelli19}, Zigbee~\cite{Bianchi19}, and Wi-Fi~\cite{He16}), ultra-wideband (UWB) is known to achieve high positioning accuracy, as it communicates on a large bandwidth that provides high distance resolution~\cite{Mazhar17}.
In addition, UWB is known to have a high signal-to-noise ratio (SNR) and penetration ability, from which more reliable and robust WP can be performed~\cite{Gezici09}.

Existing WP algorithms can be categorized into two classes: geometric methods and fingerprinting methods~\cite{Mazhar17}.
Geometric methods require each sensor to take a set of informative measurements from the exchanged signals and transfer them to the DFC.
Potential measurements include received signal strength (RSS)~\cite{Yuan23}, time of arrival (TOA)~\cite{Ni19}, time difference of arrival (TDOA)~\cite{Huang15}, and angle of arrival (AOA)~\cite{Arash20}.
Using these measurements, the DFC predicts the target location via a standard estimation algorithm (e.g., weighted least squares or gradient descent).
Fingerprinting methods, in contrast, utilize a pre-acquired set of labeled measurements (i.e., the location information is available for each measurement obtained) to feed data-driven approaches for estimating the target location~\cite{Bshara10,Subedi17}.
The labeled data can be used to compare with incoming measurements directly (e.g., through nearest-neighbor methods~\cite{Li16}) or as training data for learning a parametric classification model like a support vector machine (SVM).

While geometric methods typically involve considerably lower complexity than fingerprinting methods, the latter approaches usually lead to more accurate and more robust performance~\cite{Mazhar17}.
For example, WP using TOA measurements (a geometric method) shows low accuracy when the channel experiences non-line-of-sight (NLOS) conditions, calling for compensation techniques to recover the performance~\cite{Guvenc09,Katwe20}.
On the other hand, in order to improve accuracy and achieve robustness against varying channel conditions, fingerprinting often uses a large dimensional input feature space -- commonly, the power delay profile (PDP) -- which can lead to high complexity to carry out the training process.

With the rapid development of machine learning (ML) techniques, research on \textit{learning-based} WP has recently progressed.
Deep learning frameworks have proven to be effective solutions to various fingerprinting-based WP approaches~\cite{Fayyad20,Alhomayani20,Feng22}.
In particular, neural networks have been shown to successfully handle key tasks of WP like location estimation~\cite{Yu11}, ranging error mitigation~\cite{Wymeersch12}, and channel condition classification~\cite{Kim23}.
Moreover, various types of neural networks have been applied to solve WP problems in complex channel environments.
For example, WP algorithms using convolutional neural networks (CNN)~\cite{Nguyen20}, long short-term memory (LSTM)~\cite{Poulose20}, and gated recurrent units (GRU)~\cite{Nguyen21} have shown improved performance across different channel conditions and positioning environments.
Additionally, more recent works on learning-based WP have adopted new learning mechanisms (e.g., model-agnostic meta-learning~\cite{Gao23} and knowledge transfer~\cite{Li21}) to improve the performance further.

Although these works have shown promising results and significantly contributed to deep learning-based WP, processing high-dimensional features as is often required can become a limiting factor for many mobile applications.
For one, in PDP-based approaches, this data must be measured and collected for each positioning instance, which naturally imposes a large bandwidth and/or a long latency on the sensor network.
Also, neural networks with high-dimensional features may require high computational power (i.e., costly hardware) to support the fast positioning rates~\cite{Fontaine20}.
These operational constraints can be undesirable especially for devices or machines in which both latency and cost are critical factors.
While there exist some works that utilize low-dimensional feature data (e.g., TOA/RSS-based WP via neural networks and a linear estimator~\cite{Zheng23}), their performance is still heavily impacted by channel conditions, which may require additional tasks like ranging error detection~\cite{Kim23}.

To address this issue, metaheuristic-based feature selection methods have been recently proposed in wireless positioning~\cite{Roy21,Lalama22}. 
In some of these works, the feature set is refined by an access point selection step, conducted via e.g., binary particle swarm optimization~\cite{Panja22} or genetic algorithm~\cite{Zhang21}.
Moreover, the work in~\cite{Vijayanand20} adopted whale optimization algorithm~\cite{Nadimi23} to determine a set of effective features for intrusion detection.
While these metaheuristic approaches show effective performance in feature selection, the algorithms in general require careful fine-tuning of their feature size and search space, which we aim to eliminate with our method.
Also, we aim to incorporate more wireless-specific modeling into our approach (e.g., leveraging the channel properties) to exploit the wireless positioning setup.

In this work, we consider WP conducted in a mobile environment and propose a neural network methodology that relies on a \textit{minimum description feature set} to enjoy an improved performance-complexity tradeoff.
Instead of using the full PDP, we propose using only the largest power measurements and their temporal locations to generate a low-dimensional feature set.
We specifically design an architecture called the positioning neural network (P-NN), which takes these features as sparse image and measurement matrices and processes them using a set of convolutional, self-attention, and fully-connected layers.
We also develop a method for adaptively selecting the size of our feature set to keep the performance robust across diverse channel conditions.
The method adopts the principle of model order selection and leverages the criterion formulated with the log-likelihood, acquisition probability, and Kullback-Leibler (KL) divergence.
Numerical results show that our P-NN can provide classification accuracies and robustness matching more computationally expensive baselines and thus achieve better performance-complexity tradeoff.

\section{Summary of Contributions}\label{sec:contributions}

We summarize our contributions as follows:
\begin{itemize}
    \item For learning-based WP, we propose a new set of minimum description features consisting of the largest power measurements and their temporal locations. Compared to using the entire PDP set, the proposed feature set has significantly reduced dimensions, yet still provides information needed for accurate WP.
    
    \item We design a neural network called P-NN which takes our proposed feature set as an input to perform the WP classification task. The network adopts a multi-channel approach and transforms the input into measurement matrices and a sparse image and processes them via convolutional layers, a self-attention layer, and fully-connected layers to improve the efficiency in information extraction.
    
    \item We develop a method of adaptively selecting the size of our feature space that ensures robust performance across varying channel conditions. Our method adopts a model order selection approach, where the cost function is formulated based on log-likelihood, information acquisition probability, and KL divergence metrics to evaluate the features from both information-theoretic and classification capability perspectives.
    
    \item We analyze and characterize the behavior of the log-likelihood function for model order selection. We show that maximizing the log-likelihood function leads to a desirable size for our feature set in high SNR regimes.
    
    \item We provide a set of numerical experiments to evaluate P-NN.
    The results show that our feature set provides competitive (or better) performance against the PDP-based baselines in high (or low) SNR regimes, and thus achieves a desirable performance-complexity tradeoff.
\end{itemize}

The rest of the paper is organized as follows.
We first provide details of the WP system model in Sec.~\ref{sec:system_model}.
After describing our proposed features and the architecture of P-NN in Sec.~\ref{sec:positioning_network}, we present our feature size selection method in Sec.~\ref{sec:size_selection}.
We conduct exhaustive numerical experiments and discuss their results in Sec.~\ref{sec:numerical}, and then  Sec.~\ref{sec:conclusion} concludes the paper. 

\section{System Model}\label{sec:system_model}

\begin{figure}[!t]
    \centering
    \begin{subfigure}[!h]{0.52\linewidth}
        \centering
        \includegraphics[width=1\linewidth]{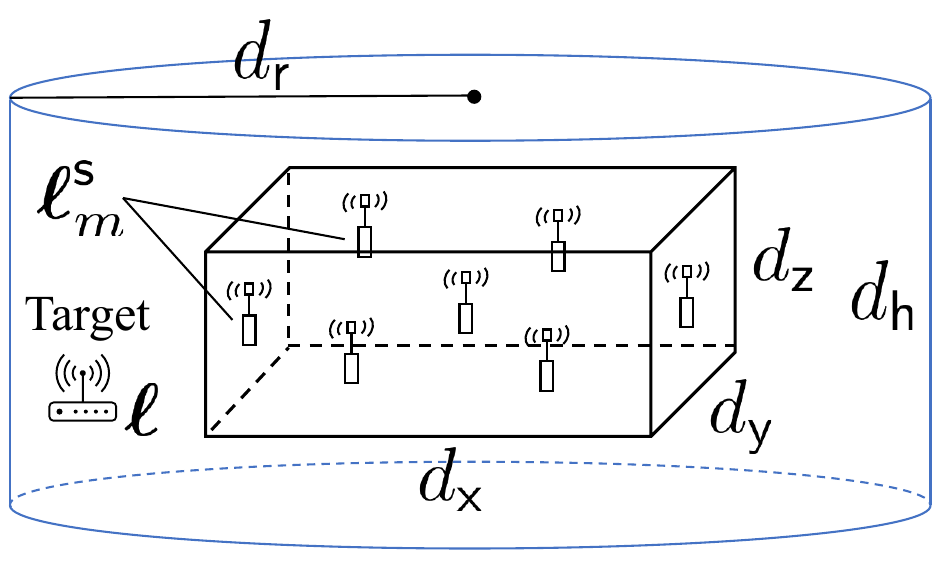}
        \caption{Layout of sensor and target spaces.}
        \label{fig:system_layout}
    \end{subfigure}
    \begin{subfigure}[!h]{0.44\linewidth}
        \centering
        \includegraphics[width=0.95\linewidth]{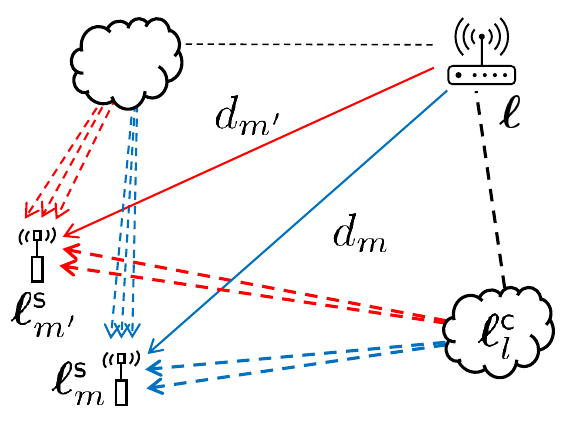}
        \caption{Channel propagation with clusters.}
        \label{fig:channel}
    \end{subfigure}
    \caption{Visual illustrations of the geographical layout of positioning spaces (left) and the channel propagation (right).}
    \vspace{-2mm}
\end{figure}

We consider the geographical layout of our WP scenario in Fig.~\ref{fig:system_layout}.
$M$ single-antenna UWB sensors are placed in a rectangular sensor space defined by the length parameters $d_\mathsf{x}$, $d_\mathsf{y}$, and $d_\mathsf{z}$. 
We use $\boldsymbol{\ell}^\mathsf{s}_m=[x^\mathsf{s}_m, y^\mathsf{s}_m, z^\mathsf{s}_m]^\top$ to denote the location of sensor $m\in\{0,1,\ldots,M-1\}$.
We aim to localize a target positioned outside the sensor space but inside a cylindrical target space defined by the radius $d_\mathsf{r}$ and height $d_\mathsf{h}$.
We assume that both the sensor and target spaces are centered at $(0,0,0)$ where $d_\mathsf{h} > d_\mathsf{z}$ and $d_\mathsf{r}^2 > (\frac{d_\mathsf{x}}{2})^2+(\frac{d_\mathsf{y}}{2})^2$ so that the entire sensor space is placed inside the target space.
Note that we specifically assume the positioning layout in Fig.~\ref{fig:system_layout} to consider WP conducted in a mobile environment (e.g., WP performed by vehicles, drones, etc.), where the sensors are relatively clustered in center and the target of interest is in general located outward.

\begin{figure}[!t]
    \centering
    \includegraphics[width=0.98\linewidth]{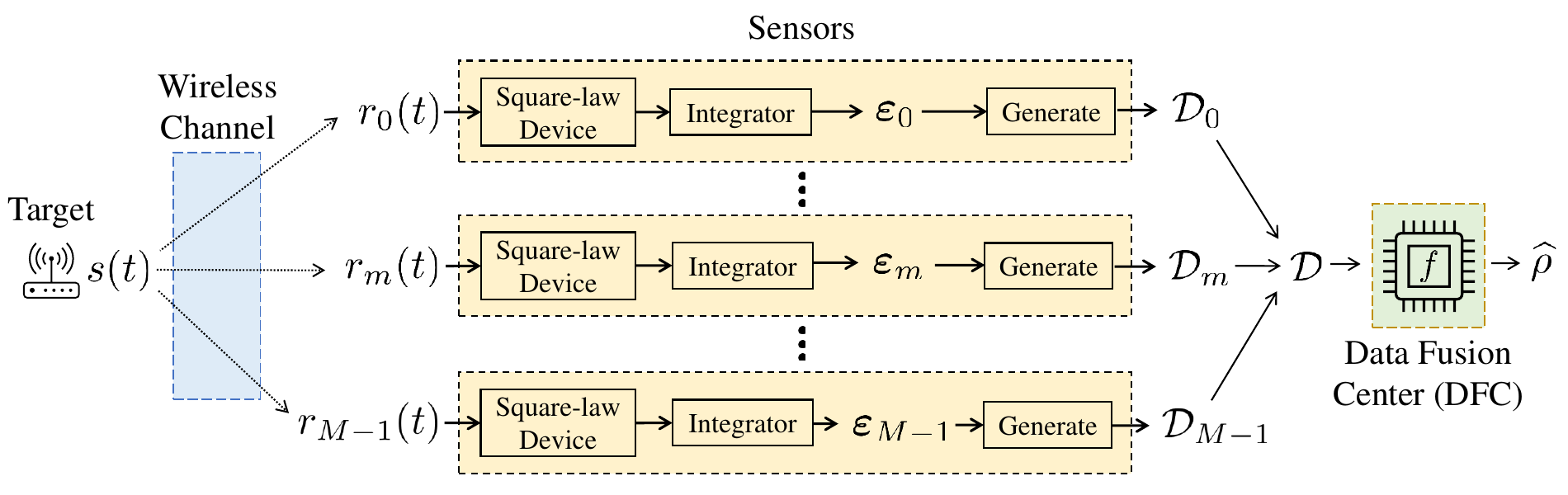}
    \caption{An overall diagram on wireless positioning using UWB sensors. Each sensor uses an energy detector for the power measurement.}
    \label{fig:system_model}
\end{figure}

The overall procedure of WP that we consider is illustrated in Fig.~\ref{fig:system_model}.
Suppose that a target located at $\boldsymbol{\ell}=[x,y,z]^\top$ transmits a radio impulse signal of duration $T_\mathsf{s}$ that is known to both the target and sensors~\cite{UWB20}.
After going through a bandpass filter of bandwidth $W$ to remove the out-of-band noise, the baseband signal received by sensor $m$ can be expressed as
\begin{align}
    r_{m}(t)=\sum^{L}_{l=0}\sum^{K_l-1}_{k=0}&a_{m,l,k}e^{j\phi_{m,l,k}}\hspace{1mm}\times \nonumber \\ & s\Big(t-\frac{d_m}{c}-T_{m,l}-\tau_{m,l,k}\Big)+\,w_{m}(t),
    \label{eq:received_signal}
\end{align}
where $L+1$ is the number of propagation paths, and $K_l$ is the number of rays existing in each path $l$~\cite{Molisch04}.
Here, we use $l=0$ to refer to the line-of-sight (LOS) path and $l=1,2,\ldots,L$ to index $L$ NLOS paths.
In~\eqref{eq:received_signal}, we use $s(t)$ to denote the lowpass equivalent representation of the transmitted impulse.
We use $a_{m,l,k}e^{j\phi_{m,l,k}}$ to denote the complex channel gain, where $a_{m,l,k}$ and $\phi_{m,l,k}$ are the weight and uniformly distributed phase, respectively.
We assume that the channel weight $a_{m,l,k}$ follows a Nakagami distribution of Nakagami factor $\mu_{m,l,k}$ and mean-square value $\Omega_{m,l,k}$.
As in~\cite{Molisch04}, we assume that the Nakagami factor $\mu_{m,l,k}$ follows a log-normal distribution of mean $\overline{\mu}$ and variance $\widetilde{\mu}$, i.e., $\ln(\mu_{m,l,k})\sim\mathcal{N}(\overline{\mu},\widetilde{\mu})$.
The term $w_m(t)$ represents zero-mean complex Gaussian noise of variance $\sigma^2_{\mathsf{n},m}$, i.e., $w_m(t)\sim\mathcal{CN}(0,\sigma^2_{\mathsf{n},m})$.

In the following, we describe the delay parameters of~\eqref{eq:received_signal}.
A visual illustration of our channel model is provided in Fig.~\ref{fig:channel}.
Let us define $d_{m}=\|\boldsymbol{\ell}^{\mathsf{s}}_m-\boldsymbol{\ell}\|_2$ as the Euclidean distance between the target and sensor~$m$.
Then, with $c$ being the speed of light constant, ${d_m}/{c}$ represents the TOA of the LOS path.
We use $T_{m,l}$ to denote the relative delay of path $l$ with respect to the LOS path, which is expressed as
\begin{equation}
    T_{m,l} =
    \begin{cases}
        0\quad\quad\quad\quad\quad\quad\quad\quad\quad\text{if }l=0, \\
        \frac{\|\boldsymbol{\ell}^{\mathsf{c}}_l-\boldsymbol{\ell}\|_2+\|\boldsymbol{\ell}^{\mathsf{s}}_m-\boldsymbol{\ell}^{\mathsf{c}}_l\|_2-d_m}{c}\;\;\text{if }l > 0,
    \end{cases}
    \label{eq:path_delay}
\end{equation}
where $\boldsymbol{\ell}^{\mathsf{c}}_l=[x^\mathsf{c}_l, y^\mathsf{c}_l, z^\mathsf{c}_l]^\top$ is the location of cluster that imposes path $l\in\{1,\ldots,L\}$.
The term $\tau_{m,l,k}$ denotes the relative delay of ray $k$ with respect to $T_{m,l}$, where $k$ is indexed in ascending order, i.e., $\tau_{m,l,k}$ increases with $k$ for given $m$ and $l$.
Hence, $\tau_{m,l,0}=0$ for all sensors and paths.
For $k>0$, we assume each ray follows the distribution of density function $p(\tau_{m,l,k}\vert \tau_{m,l,k-1})=\kappa e^{-\kappa(\tau_{m,l,k}-\tau_{m,l,k-1})}$, where $\kappa$ is the ray arrival rate~\cite{Molisch04}.
Based on the parameters defined above, the TOA of each existing channel ray is expressed as $\frac{d_m}{c}+T_{m,l}+\tau_{m,l,k}$.

We now provide the details of the channel fading model.
First, we define $\beta_{m,0,0}$ to be the pathloss of the LOS path channel between the target and sensor $m$, the expression of which is given as~\cite{Molisch04}
\begin{equation}
    \beta_{m,0,0}=\mathbb{E}[a_{m,0,0}^2]=S^{\mathsf{s}}_{m} \overline{P}_{m}\left(d_{m}\,/\,\overline{d}_{m}\right)^{-\xi},
    \label{eq:LOS_pathloss}
\end{equation}
where $\overline{P}_{m}$, $\overline{d}_{m}$, $\xi$ are the reference power, reference distance, and pathloss exponent, respectively.
$S^{\mathsf{s}}_{m}$ represents the random shadowing that follows a zero-mean log-normal distribution with variance $\sigma^2_{\mathsf{s}}$, i.e., $\ln\big(S^{\mathsf{s}}_m\big)\sim\mathcal{N}(0,\sigma^2_{\mathsf{s}})$.
The pathloss of the non line-of-sight (NLOS) path channels, denoted by $\beta_{m,l,k}$ for $l>0$ and $k>0$, is expressed as~\cite{Molisch04}
\begin{equation}
    \beta_{m,l,k}=\mathbb{E}[a_{m,l,k}^2]=S^{\mathsf{c}}_{l}\beta_{m,0,0}e^{-\frac{T_{m,l}}{\Gamma}} e^{-\frac{\tau_{m,l,k}}{\gamma}},
    \label{eq:NLOS_pathloss}
\end{equation}
where $\Gamma$ and $\gamma$ are the cluster and ray decaying constants, respectively.
The term $S^{\mathsf{c}}_{l}$ denotes the cluster shadowing that follows zero-mean log-normal distribution with variance $\sigma^2_{\mathsf{c}}$, i.e., $\ln \big(S^{\mathsf{c}}_l\big)\sim\mathcal{N}(0,\sigma^2_{\mathsf{c}})$.
With~\eqref{eq:LOS_pathloss} and~\eqref{eq:NLOS_pathloss}, each pathloss becomes strongly dependent on the channel propagation distance, which allows the channel paths to convey spatial correlation.
To make the channel fading reflect the pathloss, we set $\Omega_{m,l,k}=\beta_{m,l,k}, \forall m,l,k$.

We assume that the signal $s(t)$ is transmitted within a frame of duration $T_\mathsf{f}$ such that $ T_\mathsf{f} > \max_{m,l,k}(\frac{d_m}{c}+T_{m,l}+\tau_{m,l,k})$ (i.e., the frame has a guard period).
This ensures that each sensor safely captures $r_m(t)$ and avoids inter-signal interference.
In each sensor, the received frame is processed by an energy detector\footnote{Instead of applying a matched filter~\cite{Dardari08}, which requires at least the Nyquist sampling rate, and, thus, imposes a significant increase in the implementation complexity, our work adopts a low-complexity energy detector that can operate on sub-Nyquist rates to consider mobile applications with low-cost sensors. We will numerically evaluate the difference between these schemes in Section~\ref{sec:numerical}.} that consists of a square-law device and an integrator~\cite{Dardari08,Giorgetti13}.
For integration, the frame is broken down to $N_\mathsf{b}=\lfloor\frac{T_\mathsf{f}}{T_\mathsf{g}}\rfloor$ temporal bins, where $T_\mathsf{g}$ is the integration period, and the power contained in each temporal bin $n\in\{0,1,\ldots,N_\mathsf{b}-1\}$ of sensor $m$ is measured as
\begin{equation}
    \varepsilon_{m,n} =\frac{1}{2}\int^{(n+1)T_\mathsf{g}}_{nT_\mathsf{g}}\left\vert r_m(t)\right\vert^2dt.
    \label{eq:bin_power_continous}
\end{equation}
Now, we define the instant PDP vector measured at sensor $m$ as $\boldsymbol{\varepsilon}_m = [\varepsilon_{m,0},\varepsilon_{m,1},\ldots,\varepsilon_{m,N_\mathsf{b}-1}]^\top$.
For a signal of bandwidth $W$,~\eqref{eq:bin_power_continous} can be written as~\cite{Urkowitz67,Dardari08}
\begin{equation}
    \varepsilon_{m,n} = \frac{1}{2W}\sum^{2WT_\mathsf{g}-1}_{i=0}\left\vert r_m\left(nT_\mathsf{g}+\frac{i}{2W}\right)\right\vert^2.
    \label{eq:bin_power}
\end{equation}

\begin{figure}[!t]
    \centering
    \begin{subfigure}[!h]{0.49\linewidth}
        \centering
        \includegraphics[width=1\linewidth]{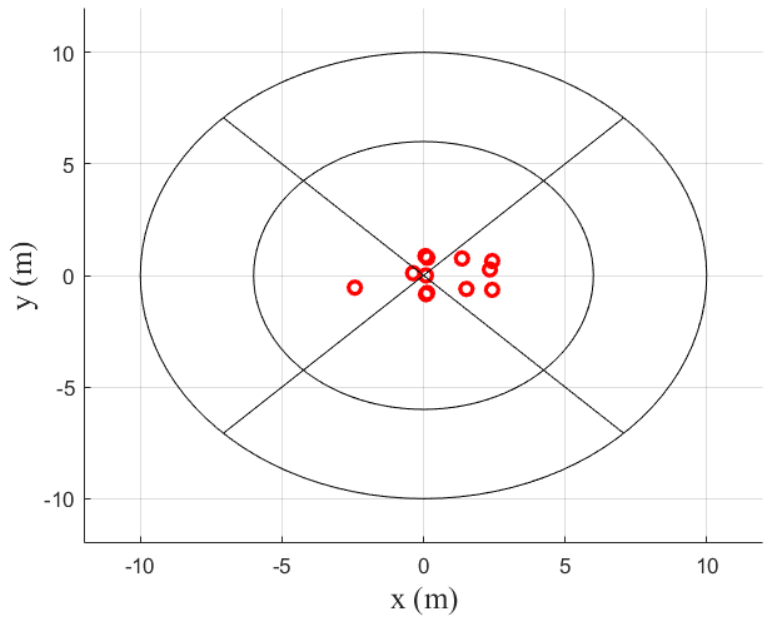}
        \label{fig:4x2}
    \end{subfigure}
    \begin{subfigure}[!h]{0.49\linewidth}
        \centering
        \includegraphics[width=1\linewidth]{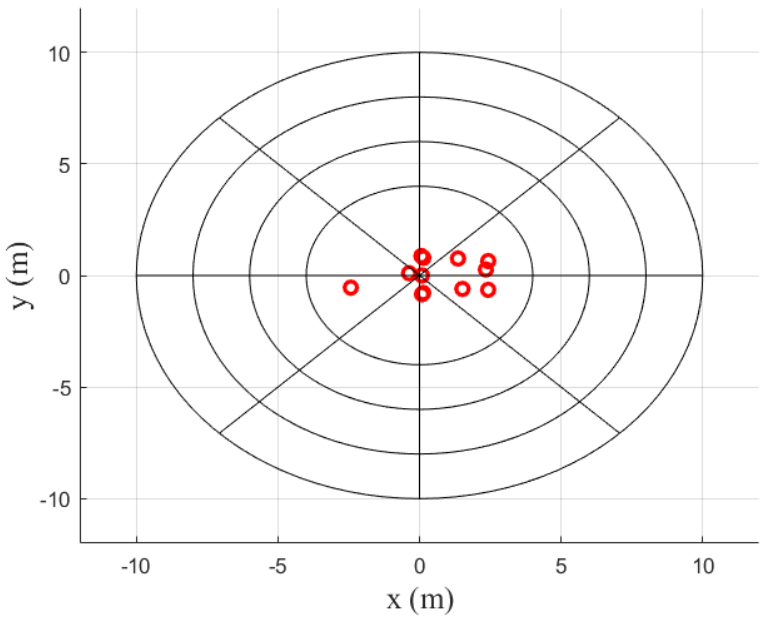}
        \label{fig:8x4}
    \end{subfigure}
    \caption{Zone layouts with $N_\mathsf{z}=8$ (left) and $N_\mathsf{z}=32$ (right). Zones are created using radius and angle for practical outward positioning settings. Red circles indicate sensor positions.}
    \label{fig:zone_layouts}
    \vspace{-2mm}
\end{figure}

Each sensor $m$ generates a data set $\mathcal{D}_m$ from $\boldsymbol{\varepsilon}_m$ and transfers it to the DFC.
Using the collected set $\mathcal{D}=\{\mathcal{D}_m\}_{m=0}^{M-1}$, the DFC estimates the target~location.
In this work, we frame our WP as an $N_\mathsf{z}$-zone classification task.
Example layouts for $N_\mathsf{z}=8$ and $N_\mathsf{z}=32$ are provided in Fig.~\ref{fig:zone_layouts}, where the zones are created using radii and angles for practical mobile application settings.
We pursue the zone classification task for the following reasons.
First, rather than coordinate-level localization, positioning via $N_\mathsf{z}$ spatial zones is often sufficient in many vehicular operations, as the value of $N_\mathsf{z}$ can be adjusted to satisfy the positioning sensitivity and resolution.
Second, it is more difficult to obtain coordinate-labeled training data than zone-labeled data.
Hence, we define our positioning task using a function $f:\mathcal{D}\rightarrow\widehat{\rho}$, where $\widehat{\rho}\in\{0,1,\ldots,N_\mathsf{z}-1\}$ is the output indicating one of the $N_\mathsf{z}$ zones.
Letting $\rho\in\{0,1,\ldots,N_\mathsf{z}-1\}$ denote the zone in which the target is truly located,
the target is correctly positioned if $\widehat{\rho}=\rho$.

\section{Positioning Neural Network (P-NN)}\label{sec:positioning_network}

In this section, we provide implementation details of our P-NN, which executes the estimation function $f$ of the DFC in Fig.~\ref{fig:system_model}.
In Sec.~\ref{ssec:features}, we present our proposed set of minimum description features and provide the motivation.
Then, we describe the architecture of P-NN in Sec.~\ref{ssec:architecture}.
In Sec.~\ref{ssec:operation}, we explain the training and testing steps of our P-NN. 

\subsection{Features of Minimum Description Length}\label{ssec:features}

Many deep learning-based WP algorithms directly use full PDP data (i.e., $\mathcal{D}=\{\boldsymbol{\varepsilon}_m\}_{m=0}^{M-1}$) to achieve high positioning accuracy and robust performance.
Processing such high-dimensional features, however, often increases the operation requirement (e.g., bandwidth, memory, and power) since the data must be measured, collected, and processed by every positioning instance.
This can be prohibitive especially for mobile applications where the operational resources are fundamentally limited.
Here, we follow the principle of minimum description length (MDL)~\cite{Wax85} that the best model for describing data is one with the smallest size, and propose to use only a small number of the largest power measurements and their temporal locations.

Suppose that each sensor $m$ receives the signal $r_m(t)$ and measures the PDP vector $\boldsymbol{\varepsilon}_m$ of size $N_\mathsf{b}$.
The elements of $\boldsymbol{\varepsilon}_m$ are then sorted in the descending order to yield
\begin{equation}
    \boldsymbol{\varepsilon}_m^{\text{ord}}=[\varepsilon_{m,0}^{\text{ord}},\varepsilon_{m,1}^{\text{ord}},\ldots,\varepsilon_{m,N_\mathsf{b}-1}^{\text{ord}}]^\top,
    \nonumber
\end{equation}
which satisfies $\varepsilon_{m,0}^{\text{ord}}\geq\varepsilon_{m,1}^{\text{ord}}\geq\ldots\geq\varepsilon_{m,N_\mathsf{b}-1}^{\text{ord}}$.
The sensor also acquires the index vector
\begin{equation}
    \boldsymbol{b}_m^{\text{ord}}=[b_{m,0}^{\text{ord}},b_{m,1}^{\text{ord}},\ldots,b_{m,N_\mathsf{b}-1}^{\text{ord}}]^\top,
    \nonumber
\end{equation}
where $b_{m,n}^{\text{ord}}$ is the index of $\boldsymbol{\varepsilon}_m$ pointing to the entry value $\varepsilon_{m,n}^{\text{ord}}$ (i.e., $b_{m,n}^{\text{ord}}$ indicates the temporal location in $\boldsymbol{\varepsilon}_m$ where the $n$-th largest power has been measured).
The sensor then takes the first $F$ entries of both $\boldsymbol{\varepsilon}_m^{\text{ord}}$ and $\boldsymbol{b}_m^{\text{ord}}$ to generate $\mathcal{D}_m = \{\varepsilon_{m,0}^{\text{ord}},\ldots,\varepsilon_{m,F-1}^{\text{ord}},b_{m,0}^{\text{ord}},\ldots,b_{m,F-1}^{\text{ord}}\}$ of size $2F$ and transfers it to the DFC.
As a result, the feature set $\mathcal{D}$ of size $2FM$ is collected at the DFC.

The key motivation for our feature set is an assumption that information needed for accurate WP is more likely present in the temporal bins of the largest powers.
Effective TOA estimation algorithms, e.g.,~\cite{Dardari08,Giorgetti13}, are based on this assumption and use the power threshold to detect signals of significant power.
In geometric WP algorithms, both RSS and TOA measurements become useful information for conducting WP~\cite{Mazhar17}.
Therefore, we use both $\boldsymbol{\varepsilon}_m^{\text{ord}}$ and $\boldsymbol{b}_m^{\text{ord}}$, which respectively represent RSS and TOA, to generate our feature set.

Using the full PDP is informative because the entire $N_\mathsf{b}M$ measurements are perceived as an image for neural networks to train and learn.
By representing the PDP in a form of image, the information needed to perform WP (e.g., the power and delay of signals received over multiple channel propagation paths) is converted to the spatial correlation across the image.
However, if only a small fraction of $N_\mathsf{b}$ measurements actually convey useful information, it is more beneficial to process those measurements only.
Nevertheless, taking the largest powers from $N_\mathsf{b}$ measurements (i.e., the first $F$ entries of $\boldsymbol{\varepsilon}_m^{\text{ord}}$) can essentially lose information within the time domain.
Hence, we directly include the temporal information (i.e., the first $F$ entries of $\boldsymbol{b}_m^{\text{ord}}$) in our feature set.

Compared to having a PDP of size $N_\mathsf{b}$, using our feature set reduces the dimension by a factor of $\frac{2F}{N_\mathsf{b}}$ (e.g., $F=5$~and~$N_\mathsf{b}=100$ yield the size reduction by $\frac{1}{10}$).
Since deep learning algorithms (e.g., CNNs with per-layer complexities that quadratically increase with feature dimensions~\cite{Vaswani17}) typically involve large data to be stored, transferred, and/or processed, reduction in feature dimensions can result in benefits such as low storage, small bandwidth, and low computational complexity.

\begin{figure}[!t]
    \centering
    \includegraphics[width=1\linewidth]{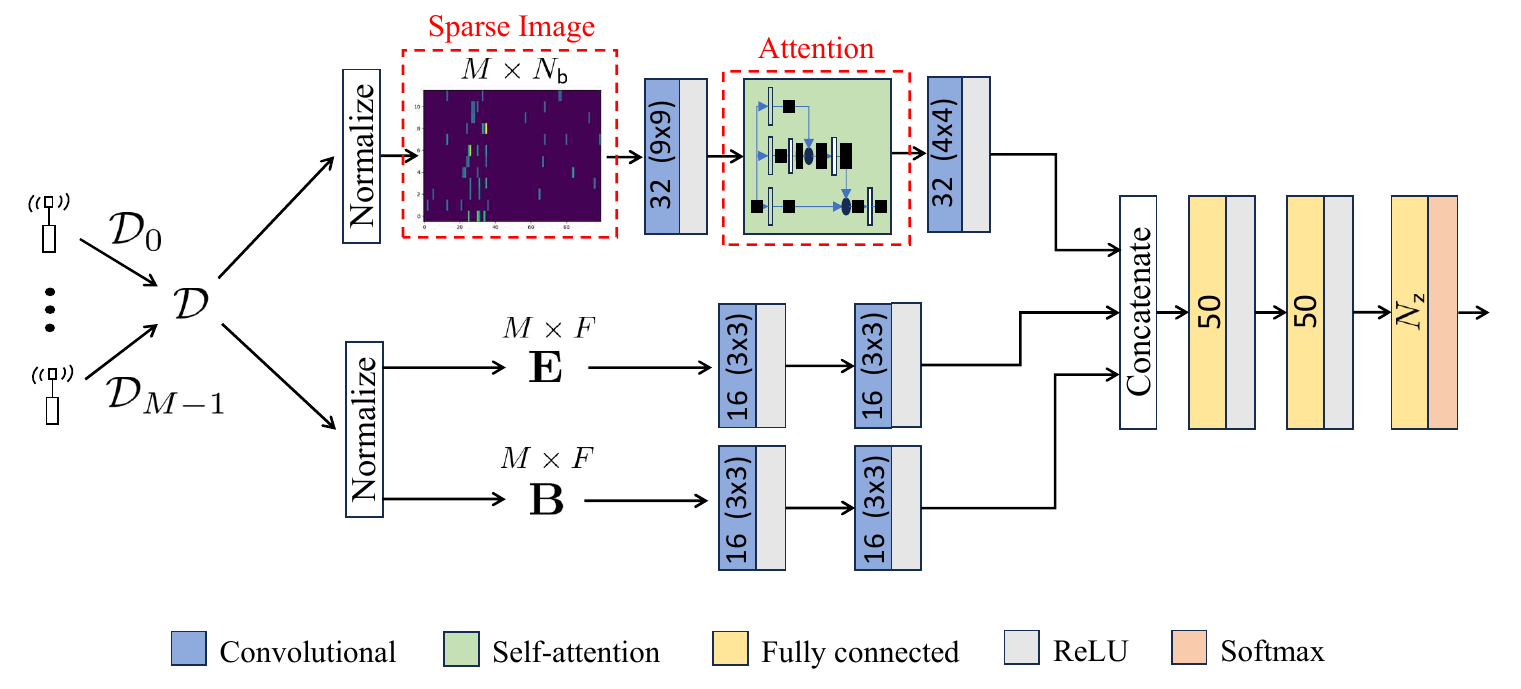}
    \caption{Architecture of our positioning neural network (P-NN). The feature set $\mathcal{D}$ is transformed into (i) a sparse image and (ii) a pair of measurement matrices, each of which goes through a different set of layers. The separately processed data sources are concatenated for combined processing.}
    \label{fig:NN_structure}
    \vspace{-2mm}
\end{figure}

\subsection{Network Architecture}\label{ssec:architecture}

Here, we provide the architectural details of our P-NN, which takes $\mathcal{D}$ as an input and outputs $\widehat{\rho}$ for the classification result.
The overall architecture of P-NN is illustrated in Fig.~\ref{fig:NN_structure}.
We design our network to take a set of inputs that are differently generated from $\mathcal{D}$.
Our P-NN first processes these inputs individually and then combines them for joint processing.
Note that such an architecture is based on the multi-channel approach, where input features are processed by several different paths to increase the information extraction capability.
In what follows, we describe the three major components of this architecture.

\subsubsection{Attention-aided spatial processing on a sparse image}\label{sssec:spatial}

\begin{figure}[!t]
    \centering
    \begin{subfigure}[!h]{0.49\linewidth}
        \centering
        \includegraphics[width=1\linewidth]{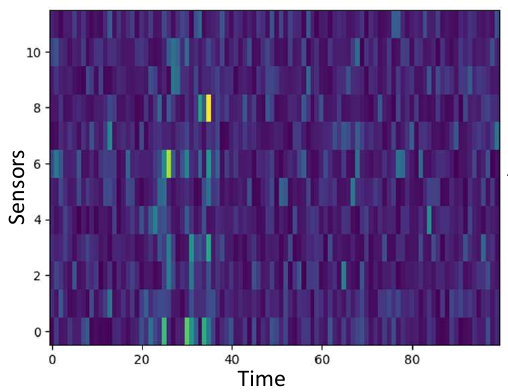}
        \label{fig:PDP_orginal}
    \end{subfigure}
    \begin{subfigure}[!h]{0.49\linewidth}
        \centering
        \includegraphics[width=1\linewidth]{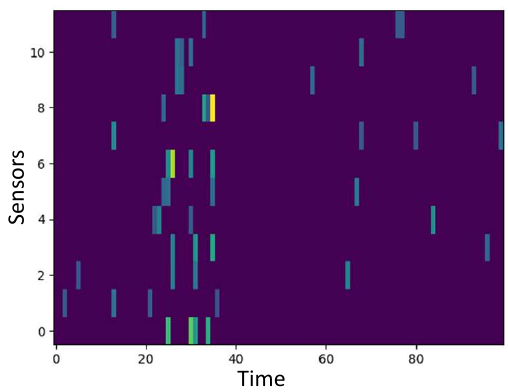}
        \label{fig:PDP_sparse}
    \end{subfigure}
    \caption{Original PDP image generated using $N_\mathsf{b}M$ measurements (left) and sparse PDP image generated using $2FM$ measurements (right), where $N_\mathsf{b}=100$, $F=4$, and $M=12$.}
    \label{fig:PDPs}
    \vspace{-2mm}
\end{figure}

The input is an $M\times N_\mathsf{b}$ sparse image generated from the $FM$ largest power measurements and their temporal locations.
Note that prior to generating the image, the power and temporal measurements are first normalized by subtracting and then dividing the data with mean and standard deviation values, respectively.
Here we compute both the mean and standard deviation values from the training set.
As discussed in Sec.~\ref{ssec:features}, PDP data is often processed as an image since the location information is spatially conveyed across both the temporal bin $n$ and sensor $m$.
Hence, as in~\cite{Nguyen20}, transforming the feature into an image format and feeding it through convolutional layers, which are particularly suited for spatial processing, is expected to be an effective approach.
Our work takes a similar approach, but we only create a sparse image by placing $FM$ power measurements at their corresponding locations.
We provide a visual illustration of our sparse image in Fig.~\ref{fig:PDPs}.
From the figure, we see that each row of the sparse image has only $F$ non-zero points, where the magnitude is indicated by the distinctiveness of color.

By processing the sparse image, we can attain the following two advantages.
First, aligned with our main objective, the number of measurements needed to be collected for conducting WP is substantially reduced as compared to using the entire PDP.
Note that we still use our feature set $\mathcal{D}$ of size $2FM$ to create an image.
Second, as we generate our sparse image only using a set of large powers, the measurements from noise-only temporal bins are likely to be discarded.
This allows our neural network to concentrate only on the expressive portion of the image and avoid being trained by the noise measurements.

To process the sparse image input, we use a set of convolutional layers with rectified linear unit (ReLU) activation to capture any significant correlation in the spatial domain.
Note that the key role of convolutional layers is to spatially process a given image.
Hence, to reinforce the capability of our spatial processing, we insert an additional layer called self-attention as shown in Fig~\ref{fig:NN_structure}.
Self-attention is a layer that is designed to detect correlations present across certain parts of an input data.
Different from the convolutional layer which focuses on correlating a given image to its label, the attention layer focuses on learning the correlation among different local regions within the image.
The effectiveness of using an attention layer has been proven in computer vision and phrase recognition.
Particularly, we implement the self-attention layer introduced in~\cite{Zhang19} to create synergy with our convolutional layers.

\begin{figure}[!t]
    \centering
    \includegraphics[width=1\linewidth]{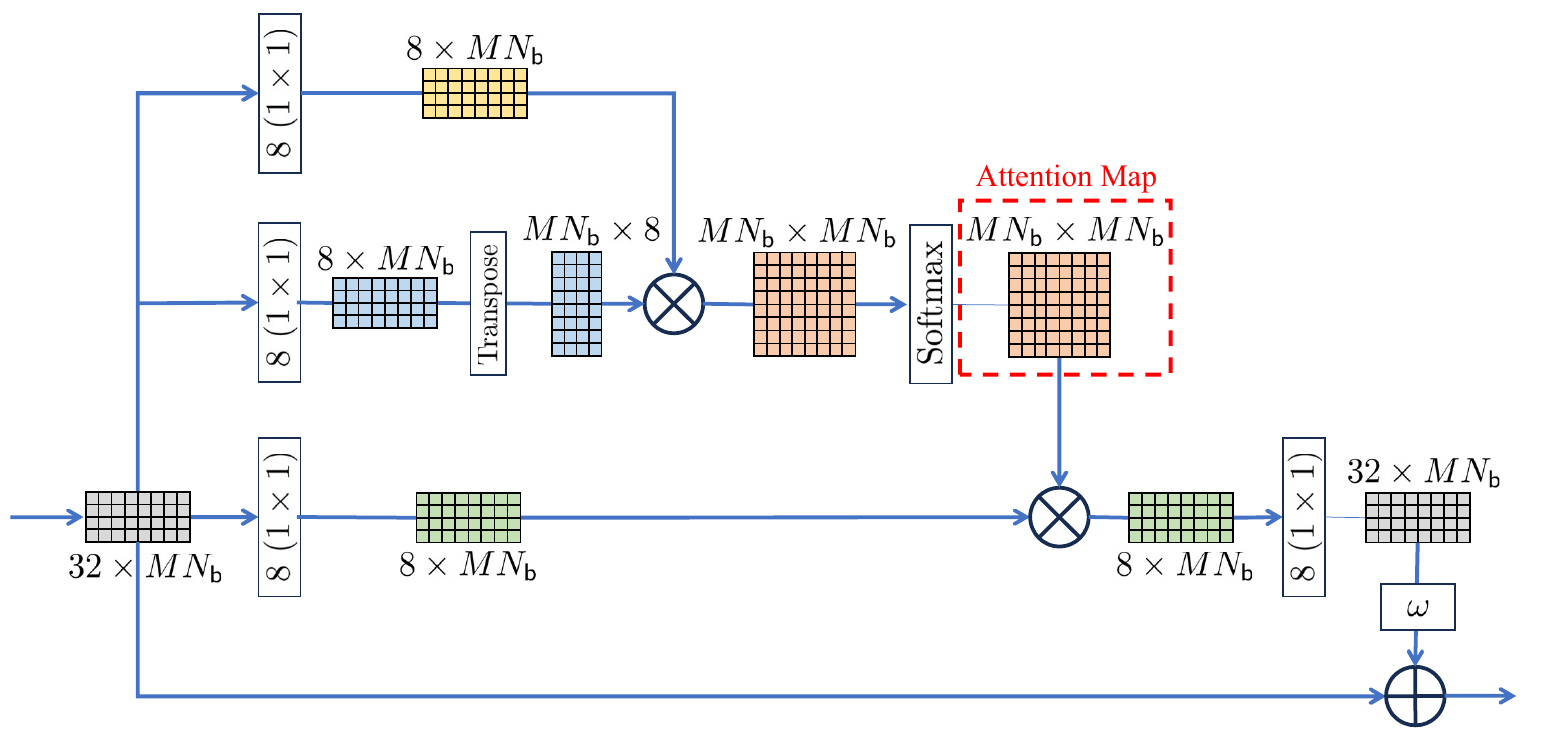}
    \caption{Structure of the implemented self-attention layer.}
    \label{fig:SA_structure}
    \vspace{-2mm}
\end{figure}

In the following, we describe the steps performed by the self-attention layer.
The structure of our self-attention layer is provided in Fig.~\ref{fig:SA_structure}.
We first reshape the input to $32\times MN_\mathsf{b}$ and process it with three individual $1\times1$ convolutional layers of eight channels to generate the components: query, key and value.
Each step here can be expressed as
$f_\mathsf{q}(\mathbf{X})=\mathbf{W}_\mathsf{q}\mathbf{X}$, $f_\mathsf{k}(\mathbf{X})=\mathbf{W}_\mathsf{k}\mathbf{X}$, and $f_\mathsf{v}(\mathbf{X})=\mathbf{W}_\mathsf{v}\mathbf{X}$,
where $\mathbf{X}$ is the $32\times MN_\mathsf{b}$ reshaped input and $\mathbf{W}_\mathsf{q}$, $\mathbf{W}_\mathsf{k}$, and $\mathbf{W}_\mathsf{v}$ are the $8 \times 32$ weight matrices corresponding for query, key, and value, respectively.

The query and key are combined via matrix multiplication and activated with the softmax function to yield the $MN_\mathsf{b} \times MN_\mathsf{b}$ attention map $\mathbf{A}$, which can be expressed as
$\mathbf{A} = f_\mathsf{sm}(f_\mathsf{q}(\mathbf{X})^\top f_\mathsf{k}(\mathbf{X}))$, 
where $f_\mathsf{sm}(\cdot)$ is the column-wise softmax operation.
Note that this attention map $\mathbf{A}$ is the key aspect of our self-attention layer as each matrix element represents the degree of attention we need to put when processing two specific regions of the image input together.
In other words, the value of $[A]_{i,j}$ indicates how much attention the model needs to give on region $i$ when it processes region $j$ of the image.

Now, the attention map $\mathbf{A}$ obtained from the query and key is multiplied with our value $f_\mathsf{v}(\mathbf{X})$ to yield an output of size $8 \times MN_\mathsf{b}$.
The output then goes through a $1\times1$ convolutional layer of 32 channels to generate a $32 \times MN_\mathsf{b}$ matrix $\mathbf{O}$, which can be expressed as
$\mathbf{O} = \mathbf{W}_\mathsf{z}f_\mathsf{v}(\mathbf{X})\mathbf{A}$, 
where $\mathbf{W}_\mathsf{z}$ is the $32 \times 8$ weight matrix for the convolutional layer.
As the last step, the matrix $\mathbf{O}$ is combined with the original input $\mathbf{X}$ using a trainable scalar weight $\omega$, i.e.,
\begin{equation}
    \mathbf{Y} = \omega\mathbf{O} + \mathbf{X},
    \label{eq:attention_update}
\end{equation}
where $\mathbf{Y}$ becomes the final output of our self-attention layer.
The weight $\omega$ is initialized as zero to make our neural network to focus on local regions first (i.e., the self-attention layer has no impact in the overall learning via $\omega=0$)~\cite{Zhang19}.
Through training, the self-attention layer gradually captures the attention and feeds it to the network via~\eqref{eq:attention_update}.

By inserting the self-attention layer for our sparse image processing, we aim to reinforce the learning ability of our P-NN.
Note that the operation of self-attention layer we implement can be simply described using linear operations of multiple weights $\mathbf{W}_\mathsf{q}$, $\mathbf{W}_\mathsf{k}$, $\mathbf{W}_\mathsf{v}$, $\mathbf{W}_\mathsf{z}$, and $\omega$.
As compared to adding a recurrent layer to the neural network for extracting attention, self-attention layer does not impose a sequential operation and provides training models that are easier to interpret~\cite{Vaswani17}.

\subsubsection{Direct processing of power and time measurement matrices}\label{sssec:direct}

For another input format, we separate the power and time measurements from $\mathcal{D}$, normalize them using the mean and standard deviation values obtained from the training data, and generate two $M\times F$ matrices
\begin{equation}
    \mathbf{E}=\hspace{-1.5mm}
    \begin{bmatrix}
	\varepsilon_{0,0}^{\text{ord}} & \hspace{-3mm}\cdots& \hspace{-3mm}\varepsilon_{0,F-1}^{\text{ord}} \\
	\vdots & \hspace{-3mm}\ddots & \hspace{-3mm}\vdots \\
	\varepsilon_{M-1,0}^{\text{ord}} & \hspace{-3mm}\cdots & \hspace{-3mm}\varepsilon_{M-1,F-1}^{\text{ord}}
    \end{bmatrix}
    \text{and }\mathbf{B}=\hspace{-1.5mm}
    \begin{bmatrix}
	b_{0,0}^{\text{ord}} & \hspace{-3mm}\cdots & \hspace{-3mm}b_{0,F-1}^{\text{ord}} \\
	\vdots & \hspace{-3mm}\ddots & \hspace{-3mm}\vdots \\
	b_{M-1,0}^{\text{ord}} & \hspace{-3mm}\cdots & \hspace{-3mm}b_{M-1,F-1}^{\text{ord}}
    \end{bmatrix}\hspace{-1mm}.
    \nonumber
\end{equation}
As shown in Fig.~\ref{fig:NN_structure}, we feed each $\mathbf{E}$ and $\mathbf{B}$ into a separate neural network first to handle the data obtained from two different domains.
Here we use two convolutional layers with ReLU activation to capture spatial correlation across both the measurements and sensors.

Recall that, in our sparse image generation, the temporal information is exploited through the $F$ largest power measurements being placed in specific locations.
Then, we rely on the learning ability of convolutional layers to successfully capture the spatial correlation.
Different from our sparse image processing, we directly feed the measurement matrices so that our network has access to the numerical values of signal powers and delays.
By doing so, we provide the network a different way to process the features and extract information.
For example, the time measurements collected in $\mathbf{B}$ can be interpreted as a set of TOA values, which is a popularly used metric in WP.

\subsubsection{Concatenation for combined processing}\label{sssec:combined}

As seen in the latter portion of our P-NN in Fig.~\ref{fig:NN_structure}, the outputs of our two separate networks (i.e., sparse image and measurement matrices processing) are flattened and concatenated to be fed to a set of two fully connected (FC) layers with ReLU activation.
The very last layer is designed with $N_\mathsf{z}$ neurons and softmax activation to output a classification vector that is directly translated to $\widehat{\rho}$.
The latter set of FC layers is to combine the information separately extracted from the sparse image, $\mathbf{E}$, and $\mathbf{B}$ and determine the output for our zone-based positioning task.

\subsection{Network Operation}\label{ssec:operation}

Since we design our WP in the supervised learning framework, an offline training phase is required for collecting the labeled dataset. 
To train our P-NN, we first acquire a training set of size $D$, where each data point indexed by $i \in \{0,1,\ldots,D-1\}$ consists of the feature set $\mathcal{D}^{(i)}=\{\mathcal{D}^{(i)}_m\}_{m=0}^{M-1}$ and the zone index $\rho_i$ for its label.
To impose unbiased learning, we obtain approximately the same amount of data points from each zone (i.e., around $\frac{D}{N_\mathsf{z}}$ data points from each zone $\rho\in\{0,1,\ldots,N_\mathsf{z}-1\}$).
The network is trained offline via Adam optimizer~\cite{Kingma14}.
During the online testing phase, the feature set $\mathcal{D}$ is obtained from the sensors in real-time and forward-fed through the neural network to determine the positioning outcome $\widehat{\rho}$.

\section{Adaptive Feature Size Selection}\label{sec:size_selection}

As discussed in Sec.~\ref{ssec:features}, the $F$ largest powers and their temporal locations are collected from each of $M$ sensors to form our feature set of size $2FM$.
Here we develop an effective strategy to adaptively determine the value of $F$ as the number of measurements to be taken by each sensor for accurate WP varies by channel conditions.
To select the value of $F$, we adopt the principle of model order selection~\cite{Akaike74} and develop a unique feature size selection method.
Model order selection allows to effectively determine the dimension or size of a model by evaluating the criterion formulated to numerically represent the objective.

In Sec.~\ref{ssec:selection_parameters}, we define three parameters that are used to evaluate the effectiveness of our feature set when the $F$ largest power measurements are considered.
In Sec.~\ref{ssec:selection_criterion}, we present our feature size selection criterion and provide an example demonstrating the selection steps.

\subsection{Parameter Definitions}\label{ssec:selection_parameters}

\subsubsection{Information coming from $F$ signal bins}\label{sssec:information}

Note that taking the $F$ largest power measurements for our feature set can be seen as assuming $F$ out of $N_\mathsf{b}$ bins to contain the signal.
Since each sensor measures the power according to~\eqref{eq:bin_power}, these $F$ signal-contained bins are assumed to follow non-central chi-square distribution~\cite{Urkowitz67}, which we approximate using central chi-square distribution of probability density function (PDF) given as~\cite{Giorgetti13}
\begin{equation}
    f(x;\psi^2,\lambda,\nu) = \left(\frac{1}{2\eta^2}\right)^{\frac{\nu}{2}}\frac{x^{\frac{\nu}{2}-1}}{\Gamma(\frac{\nu}{2})}\exp\left(-\frac{x}{2\eta^2}\right),
    \label{eq:non-central}
\end{equation}
where $\eta^2=\sqrt{\frac{2\nu\psi^4+4\psi^2\lambda+(\nu\psi^2+\lambda)^2}{\nu(2+\nu)}}$ with $\psi^2$, $\lambda$, and $\nu$ being the non-central chi-square parameters and $\Gamma(\cdot)$ is the Gamma function.
The other $N_\mathsf{b}-F$ bins are assumed to contain the noise only, and we approximate them using central chi-square distribution (i.e., we set $\lambda = 0$ in \eqref{eq:non-central}).

\begin{figure}[!t]
    \centering
    \begin{subfigure}[!h]{0.49\linewidth}
        \centering
        \includegraphics[width=1\linewidth]{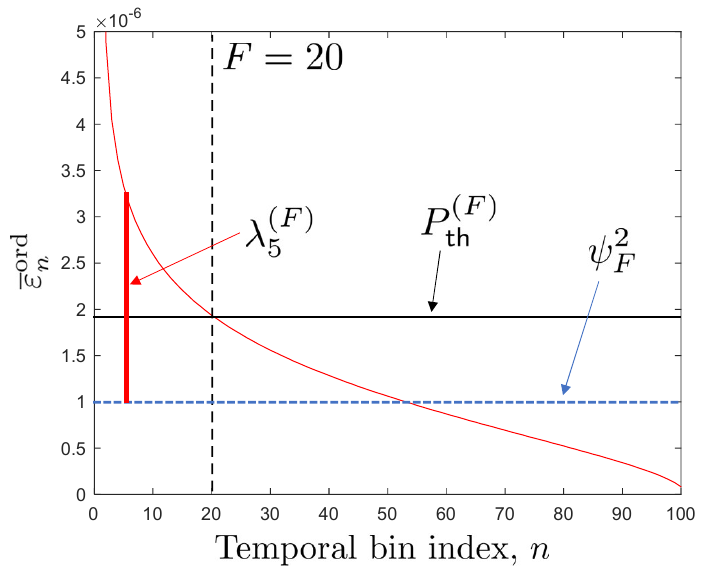}
        \caption{SNR = 5 dB}
        \label{fig:parameters_5}
    \end{subfigure}
    \begin{subfigure}[!h]{0.49\linewidth}
        \centering
        \includegraphics[width=1\linewidth]{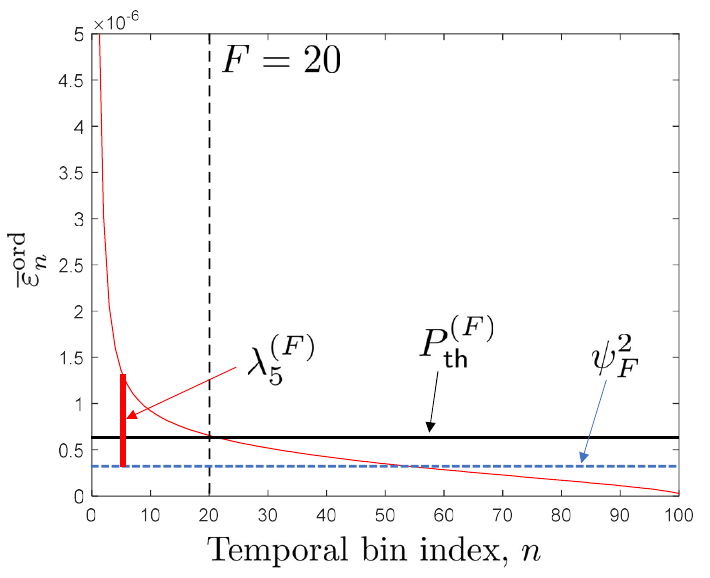}
        \caption{SNR = 10 dB}
        \label{fig:parameters_10}
    \end{subfigure}
    \caption{Visual illustration of our key measurement parameters for the case of $F=20$. Two different SNRs are considered to show the contrast.}
    \label{fig:parameters}
    \vspace{-2mm}
\end{figure}

For every data collected during the training, each sensor $m$ is supposed to find $\boldsymbol{\varepsilon}_{m}^{\text{ord}}$.
Hence, using these measurements as samples (i.e., a set of $\{\boldsymbol{\varepsilon}_{m}^{\text{ord}}\}_{m=0}^{M-1}$ that are measured to generate $D$ data points), we can compute $\overline{\boldsymbol{\varepsilon}}^\text{ord}=[\overline{\varepsilon}_{0}^{\text{ord}},\overline{\varepsilon}_{1}^{\text{ord}},\ldots,\overline{\varepsilon}_{N_\mathsf{b}-1}^{\text{ord}}]^\top$, where $\overline{\varepsilon}_{n}^{\text{ord}}$ is the power of the $n$-th largest temporal bin averaged over both the sensors and data points.
We express the joint PDF of $F$ non-central and $N_\mathsf{b}-F$ central chi-square variables using~\eqref{eq:non-central} (with appropriate values of $\lambda$) as
\begin{align}
    f(\boldsymbol{x};\psi_0^2,\ldots,&\psi_{N_\mathsf{b}-1}^2,\lambda_0,\ldots,\lambda_{F-1},\nu) \nonumber \\
    =\prod_{n=0}^{F-1}&\left(\frac{1}{2\eta_n^2}\right)^{\frac{\nu}{2}}\frac{x_n^{\frac{\nu}{2}-1}}{\Gamma(\frac{\nu}{2})}\exp\left(-\frac{x_n}{2\eta_n^2}\right) \nonumber \\
    &\times\prod_{n=F}^{N_\mathsf{b}-1}\left(\frac{1}{2\psi_n^2}\right)^{\frac{\nu}{2}}\frac{x_n^{\frac{\nu}{2}-1}}{\Gamma(\frac{\nu}{2})}\exp\left(-\frac{x_n}{2\psi_n^2}\right),
    \label{eq:joint_PDF}
\end{align}
where $\eta_n^2\hspace{-1mm}=\hspace{-1mm}\sqrt{\frac{2\nu\psi_n^4+4\psi_n^2\lambda_n+(\nu\psi_n^2+\lambda_n)^2}{\nu(2+\nu)}}$ and $\boldsymbol{x}\hspace{-1mm}=\hspace{-1mm}[x_0,\ldots,x_{N_\mathsf{b}-1}]^\top\hspace{-0.5mm}$.
From~\eqref{eq:joint_PDF}, we derive the likelihood of having $\overline{\boldsymbol{\varepsilon}}^{\text{ord}}$ as~\cite{Giorgetti13}
\begin{align}
    &\ln f(\bar{\boldsymbol{\varepsilon}}^{\text{ord}};\psi_0^2,\ldots,\psi_{N_\mathsf{b}-1}^2,\lambda_0,\ldots,\lambda_{F-1},\nu) \nonumber \\
    &=\hspace{-1mm}\sum_{n=0}^{F-1}\hspace{-1mm}-\frac{\nu}{2}\ln(2\eta^2_n)+\frac{\nu\hspace{-0.5mm}-\hspace{-0.5mm}2}{2}\ln(\overline{\varepsilon}_n^{\text{ord}})\hspace{-0.5mm}-\hspace{-0.5mm}\ln\Gamma\Big(\frac{\nu}{2}\Big)\hspace{-0.5mm}-\frac{\overline{\varepsilon}_n^{\text{ord}}}{2\eta^2_n} \nonumber \\
    &+\hspace{-1mm}\sum_{n=F}^{N_\mathsf{b}-1}\hspace{-1mm}-\frac{\nu}{2}\ln(2{\psi}^2_n)+\frac{\nu\hspace{-0.5mm}-\hspace{-0.5mm}2}{2}\ln(\overline{\varepsilon}_n^{\text{ord}})\hspace{-0.5mm}-\hspace{-0.5mm}\ln\Gamma\Big(\frac{\nu}{2}\Big)\hspace{-0.5mm}-\frac{\overline{\varepsilon}_n^{\text{ord}}}{2{\psi}^2_n}.
    \label{eq:log-likelihood}
\end{align}
Note that~\eqref{eq:log-likelihood} is characterized by $N_\mathsf{b}$ values of $\psi_n^2$, $F$ values of $\lambda_n$, and a single value of $\nu=2WT_\mathsf{g}$.
Since we do not have the knowledge of $\{\psi_n^2\}_{n=0}^{N_\mathsf{b}-1}$ and $\{\lambda_n\}_{n=0}^{F-1}$ to evaluate~\eqref{eq:log-likelihood}, we estimate each term using
\begin{align}
    \psi_F^2 &= \frac{1}{N_\mathsf{b}-F}\sum_{n=F}^{N_\mathsf{b}-1}\overline{\varepsilon}_n^\text{ord}\approx\psi_n^2,\;\;\forall n=0,\ldots,N_\mathsf{b}-1, \label{eq:psi_estimated} \\
    \lambda^{(F)}_n &= \overline{\varepsilon}_n^\text{ord} - \psi_F^2\approx\lambda_n,\;\;\forall n=0,\ldots,F-1, \label{eq:lambda_estimated}
\end{align}
where their visual illustration is provided in Fig.~\ref{fig:parameters}.
The terms~\eqref{eq:psi_estimated} and~\eqref{eq:lambda_estimated} can be respectively seen as the noise and signal powers estimated using the observations.
Using~\eqref{eq:psi_estimated} and~\eqref{eq:lambda_estimated}, we now define the estimated likelihood of having $\overline{\boldsymbol{\varepsilon}}^{\text{ord}}$ when the $F$ largest powers are taken for our feature set (i.e., $F$ bins are assumed to contain signals) as
\begin{equation}
    \hspace{-2mm}\mathsf{LL}_F = \ln f(\overline{\boldsymbol{\varepsilon}}^{\text{ord}};\psi_F^2,\ldots,\psi_F^2,\lambda_0^{(F)},\ldots,\lambda_{F-1}^{(F)},\nu).
    \label{eq:LL_F}
\end{equation}
For a given $\overline{\boldsymbol{\varepsilon}}^{\text{ord}}$, the value of~\eqref{eq:LL_F} varies by $F$, and we utilize this parameter to evaluate the expected amount of information when $F$ measurements are taken for our feature set.
Note that the log-likelihood is an effective metric popularly used for the information theoretic model order selection~\cite{Akaike74,Wax85,Giorgetti13}.

In what follows, we rationalize the usage of~\eqref{eq:LL_F} in our feature size criterion formulation by analyzing its behavior for the high SNR regime.
From~\eqref{eq:received_signal}, we have $\sum_{l=0}^{L}K_l$ independent signal paths that fall across $N_\mathsf{b}$ temporal bins, and let us define $1 \leq \tilde{F} \leq N_\mathsf{b}$ to be the number of temporal bins that actually contain these signals.
Note that $\tilde{F}$ is deterministic but unknown.
Since we desire to take the most useful information from the PDP but keep our feature dimensions as low as possible, $\tilde{F}$ intuitively becomes the ideal number of measurements for our feature size selection.

As we vary the value of $F$ to adaptively determine the feature dimension, two possible cases take place regarding the relationship between $F$ and $\tilde{F}$: (i) $F\leq\tilde{F}$ with which we select less number of measurements than desired but take a higher chance of successfully discarding noise-only measurements and (ii) $F>\tilde{F}$ where we successfully take the entire measurements from signal-contained bins but allow our features to include extra measurements that are potentially useless.

Recall that we utilize the sorted power measurement vector $\overline{\boldsymbol{\varepsilon}}^\text{ord}$ to generate our feature set.
Out of the $N_\mathsf{b}$ entries of $\overline{\boldsymbol{\varepsilon}}^\text{ord}$, $\tilde{F}$ of them contain both the signal and noise, and the rest $N_\mathsf{b}-\tilde{F}$ bins only convey the noise.
Since the power of each temporal bin is strictly dependent on the power of its components~\cite{Giorgetti13}, with high SNR, powers from the $\tilde{F}$ signal-contained bins are measured much greater than the rest and most likely placed in the first $\tilde{F}$ entries of $\overline{\boldsymbol{\varepsilon}}^\text{ord}$ after sorting.
Hence, we apply the following assumption to our analysis.
\begin{assumption}
\textup{In high SNR scenarios, the first $\tilde{F}$ entries of $\overline{\boldsymbol{\varepsilon}}^\text{ord}$ are significantly greater than the rest, and those $N_\mathsf{b}-\tilde{F}$ entries are negligibly small and approximately the same, i.e.,
\begin{equation}
    \overline{\varepsilon}^\text{ord}_0 \geq \ldots \geq \overline{\varepsilon}^\text{ord}_{\tilde{F}-1} \gg \overline{\varepsilon}^\text{ord}_{\tilde{F}}\approx\ldots\approx\overline{\varepsilon}^\text{ord}_{N_\mathsf{b}-1}.
\end{equation}}
\label{assume:high_SNR}
\vspace{-4mm}
\end{assumption}
Now, we remove expressions that are not affected by $F$ from~\eqref{eq:LL_F} for conciseness and obtain
\begin{equation}
    \widehat{\mathsf{LL}}_F =\hspace{-1mm}\sum_{n=0}^{F-1}\hspace{-1mm}-\frac{\nu}{2}\ln(2\eta^2_{F,n})-\frac{\overline{\varepsilon}^{\text{ord}}_n}{2\eta^2_{F,n}}+\hspace{-1mm}\sum_{n=F}^{N_\mathsf{b}-1}\hspace{-1mm}-\frac{\nu}{2}\ln(2{\psi}^2_F)-\frac{\overline{\varepsilon}^{\text{ord}}_n}{2{\psi}^2_F},
    \label{eq:simple_LL_F}
\end{equation}
where
\begin{equation}
    \eta_{F,n}^2=\sqrt{\frac{2\nu\psi_F^4+4\psi_F^2\lambda^{(F)}_n+(\nu\psi_F^2+\lambda^{(F)}_n)^2}{\nu(2+\nu)}}.
    \label{eq:estimated_eta}
\end{equation}
Note that, since we aim to analyze the behavior of~\eqref{eq:LL_F} in terms of our control variable $F$, \eqref{eq:simple_LL_F} becomes a sufficient expression to draw conclusions that are also applicable to~\eqref{eq:LL_F}.
Depending on the value of $F$ with respect to $\tilde{F}$, we introduce the following proposition regarding the behavior of~\eqref{eq:simple_LL_F}.
The corresponding proof is provided in Appendix~\ref{appendix:greater_F}.

\begin{proposition}
    \textup{Based on the approximation made in Assumption~\ref{assume:high_SNR}, the value of $\widehat{\mathsf{LL}}_F$, which is given by~\eqref{eq:simple_LL_F}, is a non-decreasing function of $F$ when $F\leq\tilde{F}$ and does not change with $F$ when $F>\tilde{F}$.}
    \label{prop:greater_F}
\end{proposition}

Using Proposition~\ref{prop:greater_F}, which can be also applied to~\eqref{eq:LL_F}, we claim that our log-likelihood metric~\eqref{eq:LL_F} reaches to a non-unique maximum value as $F$ approaches to $\tilde{F}$ with high SNR.
Therefore, even though we do not have the knowledge of $\tilde{F}$, maximizing~\eqref{eq:LL_F} over a given range of $F$ can lead us to the most effective decision on the size of our feature set.

\subsubsection{Information acquisition probability}\label{sssec:acquisition}

Another parameter we define is the probability of acquiring the useful information when we consider the $F$ largest power measurements.
Due to the time-varying nature of wireless channel, the power across the $N_\mathsf{b}$ temporal bins are randomly measured at each positioning instance.
In other words, despite the effort to generate our feature set using only the signal-contained bins, it is possible for the set to include measurements from the noise-only bins.
Such a case is not desirable since data with no useful information can degrade the performance of our P-NN.

Thus, for a given value of $F$, we quantify the chance of our feature set to take measurements from the signal-contained bins.
Recall that taking the $F$ largest power measurements is to assume $F$ signal-contained bins out of $N_\mathsf{b}$.
First, we define $P_\mathsf{th}^{(F)}=(\overline{\varepsilon}_{F-1}^{\text{ord}}+\overline{\varepsilon}_{F}^{\text{ord}})/2$ to be the power threshold that separates the first $F$ bins from the rest $N_\mathsf{b}-F$ bins (Fig.~\ref{fig:parameters}).
Our logic is that the feature set will likely include these signal-contained bins if their power is measured greater than $P_\mathsf{th}^{(F)}$.
Hence, using~\eqref{eq:psi_estimated} and~\eqref{eq:lambda_estimated}, we define the probability of a signal-contained bin $n\in\{0,\ldots,F-1\}$ to have the power greater than $P_\mathsf{th}^{(F)}$ as~\cite{Dardari08}
\begin{align}
    p_{n}^{(F)} & = \mathbb{P}\left\{\frac{\varepsilon^{\text{ord}}_n}{\psi^2_{F}}>\frac{P_\mathsf{th}^{(F)}}{\psi^2_{F}}\;\bigg\vert\;\frac{\lambda^{(F)}_n}{\psi^2_{F}}\right\} \nonumber \\
    & = Q_{\frac{\nu}{2}}\left(\sqrt{2(\lambda_n^{(F)}/\psi^2_{F})^2},\sqrt{2P_\mathsf{th}^{(F)}/\psi^2_{F}}\right),
    \label{eq:high_signal}
\end{align}
where $\varepsilon^\text{ord}$ is the power measured in the $n$-th largest bin, which follows a chi-square distribution of parameters $\lambda_n^{(F)}$, $\psi_{F}^2$, and $\nu$ upon assuming $F$ signal-contained bins, and $Q_{\frac{\nu}{2}}\left(\cdot,\cdot\right)$ is the $\frac{\nu}{2}$-th order Marcum Q-function~\cite{Marcum60}.
Based on~\eqref{eq:high_signal}, we define the acquisition probability of our $F$ largest powers to include the measurements from $f\in\{0,1,\ldots,F\}$ signal-contained bins as
\begin{equation}
    \mathsf{P}^{(F)}_f=\sum_{\boldsymbol{q}\in\mathcal{Q}^{(F)}_f}\prod_{i=1}^{F}(p_{i-1}^{(F)})^{\boldsymbol{q}[i]}(1-p_{i-1}^{(F)})^{(1-\boldsymbol{q}[i])},
    \label{eq:safe_capture}
\end{equation}
where $\mathcal{Q}^{(F)}_f$ is the set of all $F$-length binary vectors containing $f$ ones (i.e., $\mathcal{Q}^{(F)}_f$ considers all $\frac{F!}{f!(F-f)!}$ cases where $f$ out of $F$ bins have their power greater than $P_\mathsf{th}^{(F)}$).
The product term in~\eqref{eq:safe_capture} computes the joint probability of each case in $\mathcal{Q}^{(F)}_f$, and the summation provides the overall probability.
Note that~\eqref{eq:safe_capture} quantifies the chance of taking $f$ useful measurements when we consider the $F$ largest measurements for our feature set.
 
\subsubsection{Inter-zone Kullback-Leibler divergence}\label{sssec:KL}

Dissimilarity among the class distributions is one of the key factors that impact classification performance, and how we form our feature set directly affects this dissimilarity.
Hence, for a given value of $F$, we propose to quantify the dissimilarity across the data samples from each zone via KL divergence and use it for our feature size selection.
To evaluate KL divergence, the PDFs must be known.
Since we only have empirical measurements (i.e., training data), we take the k-nearest neighbors (KNN) density estimation approach to directly estimate the KL divergence~\cite{Perez08}.
If we subgroup the training data by each zone in terms of our feature set and denote each group using $\mathcal{D}^{\mathsf{z}}_z$ for $z\in\{0,1,\ldots,N_\mathsf{z}-1\}$, the estimated KL divergence between the zone $z$ and $z'$ using the KNN density estimation with $u$ nearest neighbors is given~by
\begin{equation}
    \hspace{-1mm}\widehat{D}_u(P_z\vert\vert P_{z'}) = \frac{F}{\vert \mathcal{D}^{\mathsf{z}}_z\vert }\hspace{-1mm}\sum_{x\in\mathcal{D}^{\mathsf{z}}_z}\hspace{-1mm}\log \frac{r_{u,z'}(x)}{r_{u,z}(x)}+\log\frac{\vert \mathcal{D}^{\mathsf{z}}_{z'}\vert }{\vert \mathcal{D}^{\mathsf{z}}_z\vert -1},
    \label{eq:KL_estimated}
\end{equation}
where $r_{u,z}(x)$ is the Euclidean distance between $x$ and its $u$-th nearest neighbor in $\mathcal{D}^{\mathsf{z}}_z$.
Now we define an empirical KL divergence upon taking the $F$ largest power measurements as
\begin{equation}
    \mathsf{KL}_F = \frac{1}{N_\mathsf{z}^2\sqrt{F}}\sum_{i=0}^{N_\mathsf{z}}\sum_{j=0}^{N_\mathsf{z}}\widehat{D}_u(P_i\vert \vert P_j),
    \label{eq:KL_mean}
\end{equation}
which we use to quantify how effectively our feature set of size $2FM$ can separate the classes.
Note that, regardless of the distributions being compared,~\eqref{eq:KL_estimated} yields a steady increase with $F$ due to the volume expression used in the KNN density estimation.
Hence, a factor of $\sqrt{F}$ is applied in~\eqref{eq:KL_mean} to account for the increase in the expected Euclidean distance across $F$.

\begin{table*}[!t]
\centering
\setlength\extrarowheight{5pt}
\setlength\tabcolsep{2pt}
\caption{Numerical values of the key parameters used in our feature size selection steps. $\psi^2_F$, $\lambda^{(F)}_n$, and $P_\mathsf{th}^{(F)}$ are in the unit of $10^{-7}$.}
\label{tb:example}
\begin{tabular}{|c|c|c|c|c|c|c|} 
    \hline
    $F$ & $3$ & $4$ & $5$ & $6$ & $7$ & $8$\tabularnewline
    \hline
    $\psi^2_F$ & 5.84 & 4.73 & 3.79 & 3.40 & 2.96 & 2.76 \tabularnewline
    \hline
    $\{\lambda^{(F)}_n\}_{n=2}^{F-1}$ & $\{11.5\}$ & $\{12.6,7.8\}$ & $\{13.6,8.7,5.7\}$ & $\{14.0,9.1,6.1,2.0\}$ & $\{14.4,9.5,6.5,2.4,1.8\}$ & $\{14.6,9.7,6.7,2.6,2.0,0.6\}$\tabularnewline
    \hline
    $\overline{\mathsf{LL}_F\hspace{-0.5mm}-\hspace{-0.5mm}\mathsf{LL}_0}$ & $0.713$ & $0.822$ & $0.919$ & $0.951$ & $0.986$ & $1$ \tabularnewline
    \hline 
    $P^{(F)}_\mathsf{th}$ & 14.93 & 10.98 & 7.41 & 5.04 & 4.04 & 3.16 \tabularnewline
    \hline
    $\{p^{(F)}_{\mathsf{s},n}\}_{n=2}^{F-1}$ & $\{0.77\}$ & $\{0.96,0.66\}$ & $\{1.00,0.93,0.65\}$ & $\{1.00,0.99,0.85,0.33\}$ & $\{1.00,1.00,0.95,0.46,0.37\}$ & $\{1.00,1.00,0.98,0.58,0.48,0.34\}$ \tabularnewline
    \hline
    $\{\mathsf{P}^{(F)}_f\}_{f=3}^{F}$ & $\{0.77\}$ & $\{0.36,0.63\}$ & $\{0.02,0.37,0.61\}$ & $\{0.00,0.10,0.61,0.28\}$ & $\{0.00,0.02,0.35,0.47,0.16\}$ & $\{0.00,0.00,0.15,0.41,0.35,0.09\}$ \tabularnewline
    \hline
    $(a)$ in~\eqref{eq:size_selection} & $0.657$ & $0.744$ & $0.842$ & $0.820$ & $0.815$ & $0.798$ \tabularnewline
    \hline
    $(b)$ in~\eqref{eq:size_selection} & $0.921$ & $0.990$ & $1$ & $0.979$ & $0.952$ & $0.926$ \tabularnewline
    \hline
\end{tabular}
\end{table*}

\begin{figure*}[!t]
\centering
\minipage{0.32\textwidth}
    \includegraphics[width=\linewidth]{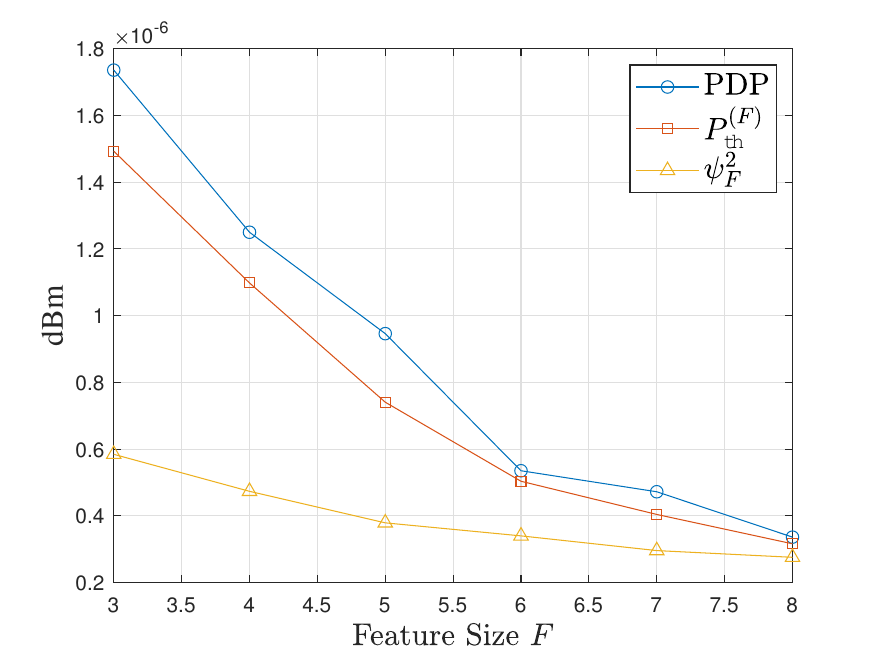}
    \caption{Numerical values of PDP, $P_\mathsf{th}^{(F)}$, and $\psi_F^2$ computed for the feature size selection example when $F\in[3,8]$.}
    \label{fig:example_1}
\endminipage\hfill
\minipage{0.325\textwidth}
    \includegraphics[width=\linewidth]{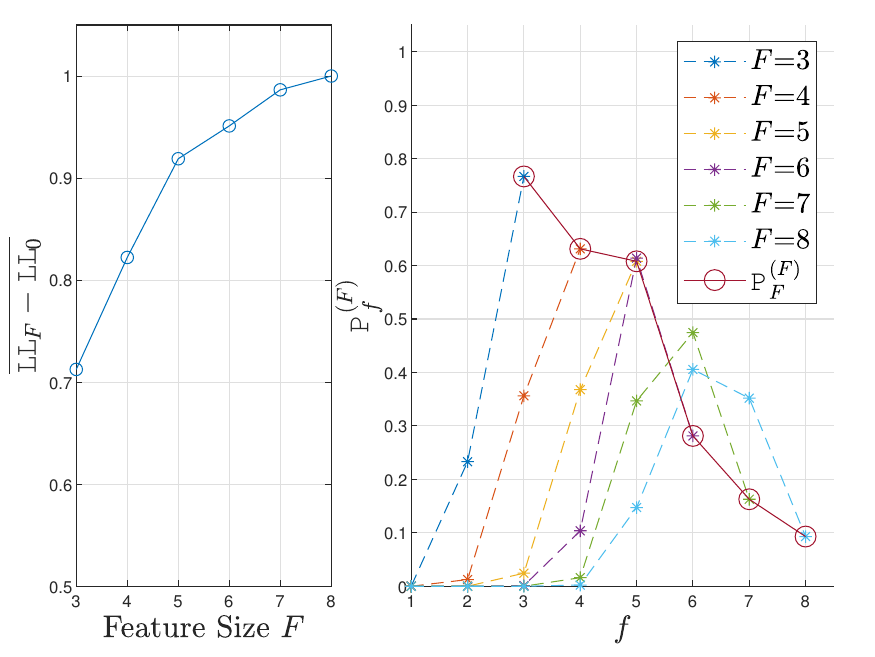}
    \caption{Numerical values computed for the feature size selection example: $\overline{\mathsf{LL}_F\hspace{-0.5mm}-\hspace{-0.5mm}\mathsf{LL}_0}$ (left) and acquisition probability $\{\mathsf{P}^{(F)}_f\}_{f=1}^{F}$  (right) for $F\in[3,8]$.}
    \label{fig:example_2}
\endminipage\hfill
\minipage{0.32\textwidth}
    \includegraphics[width=\linewidth]{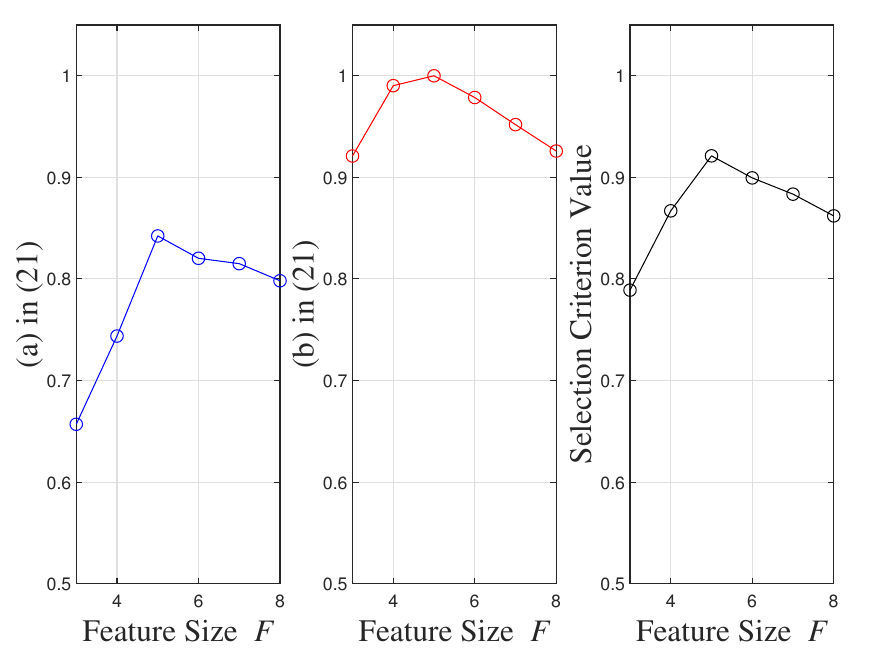}
    \caption{Numerical values computed for the feature size selection example: $(a)$ in~\eqref{eq:size_selection} (left), $(b)$ in~\eqref{eq:size_selection} (middle), and final selection criterion value (right).}
    \label{fig:example_3}
\endminipage
\end{figure*}

\subsection{Selection Criterion Formulation}\label{ssec:selection_criterion}

Using the parameters~\eqref{eq:LL_F},~\eqref{eq:safe_capture}, and~\eqref{eq:KL_mean}, we now formulate our feature size selection criterion, which is expressed as
\begin{equation}
    F^\star \hspace{-0.5mm} = \hspace{-2mm}\argmax_{F\in[F_\mathsf{min},F_\mathsf{max}]} \hspace{-1mm}\bigg(\epsilon\underbrace{\sum_{f=0}^{F}\mathsf{P}^{(F)}_f\frac{f}{F}\overline{\mathsf{LL}_F\hspace{-0.2mm}-\hspace{-0.3mm}\mathsf{LL}_0}}_{(a)} + (1-\epsilon)\hspace{-0.3mm}\underbrace{\overline{\mathsf{KL}_F}}_{(b)}\bigg)
    \label{eq:size_selection}
\end{equation}
where $\overline{(\cdot)}$ implies the normalization with respect to $\max_F(\cdot)$ and $\epsilon\in[0,1]$ is the weight parameter.
To determine $F^\star$, our feature size selection reflects two factors: the effective amount of information, i.e., $(a)$, and classification capability, i.e., $(b)$, attainable from taking the $F$ largest powers and their temporal locations.
Since the cost function is the weighted sum of $(a)$ and $(b)$, we force the range of both $(a)$ and $(b)$ to be $[0,1]$ by normalizing $\{\mathsf{LL}_F\hspace{-0.5mm}-\hspace{-0.5mm}\mathsf{LL}_0\}_{F=F_\mathsf{min}}^{F_\mathsf{max}}$ and $\{\mathsf{KL}_F\}_{F=F_\mathsf{min}}^{F_\mathsf{max}}$.
In the following, we elaborate on our choice of these cost function terms in~\eqref{eq:size_selection}.

First, we use the term $(a)$ in our cost function to reflect the effective amount of information.
Recall that $\mathsf{LL}_F$ is the log-likelihood representing the overall amount of information contained in the $F$ largest measurements.
To quantify the relative increase in information, we subtract $\mathsf{LL}_0$ from $\mathsf{LL}_F$ and normalize to compute $\overline{\mathsf{LL}_F\hspace{-0.5mm}-\hspace{-0.5mm}\mathsf{LL}_0}$.
Then, to account for the chance that only $f$ of our $F$ measurements are actually useful (i.e., the measurements are from $f$ signal-contained bins and $F-f$ noise-only bins), we weight our log-likelihood expression $\overline{\mathsf{LL}_F\hspace{-0.5mm}-\hspace{-0.5mm}\mathsf{LL}_0}$ with a factor $\frac{f}{F}$ and the acquisition probability $\mathsf{P}^{(F)}_f$.
We compute this value for each case of $f\in\{0,1,\ldots,F\}$ and sum them up to obtain the term $(a)$.
Note that, as the term reflects the likeliness of our features to include measurements from noise-only temporal bins, taking more measurements (i.e., a larger value of $F$) may not always lead to an increase in the effective amount of information.

Next, we use the term $(b)$ in our cost function to reflect the classification capability.
As explained in Sec.~\ref{sssec:KL}, the empirically estimated KL divergence in~\eqref{eq:KL_mean} serves an effective metric to quantify the dissimilarity across class distributions.
Hence, we directly adopt this parameter into our cost function to reflect the classification performance expected from utilizing the $F$ largest measurements.
Note that, unlike ~$(a)$ in~\eqref{eq:size_selection}, the term $(b)$ in our cost function relies on the statistical properties of the dataset and thus focuses on measuring the effectiveness of the dataset in differentiating the classes.

\begin{example}
We provide a numerical example of our feature size selection using the setting of $15$dB SNR and LOS condition.
For brevity, we set $N_\mathsf{b} = 10$, $[F_\mathsf{min},F_\mathsf{max}]=[3,8]$, $\nu=2$,~and $\epsilon=0.5$.
From the given setting, we assume to have obtained $\bar{\boldsymbol{\varepsilon}}^\text{ord}=[53.9,26.8,$ $17.4,12.5,9.46,5.35,4.72,3.36,2.96,2.55]\hspace{-0.5mm}\times\hspace{-0.5mm}10^{-7}$, where the first five entries contain the signal (i.e., $\tilde{F}=5$).
In Table~\ref{tb:example} and Figs.~\ref{fig:example_1},~\ref{fig:example_2}, and~\ref{fig:example_3}, we display some of the key numerical values computed for the given example.

We see that $\overline{\mathsf{LL}_F\hspace{-0.5mm}-\hspace{-0.5mm}\mathsf{LL}_0}$ shows a non-decreasing behavior in $F$ (the left plot of Fig.~\ref{fig:example_2}), which supports our Proposition~\ref{prop:greater_F}.
Note that the increase in $\overline{\mathsf{LL}_F\hspace{-0.5mm}-\hspace{-0.5mm}\mathsf{LL}_0}$ is more pronounced for $F\leq\tilde{F}$ and relatively diminished for $F>\tilde{F}$.
This implies that $\overline{\mathsf{LL}_F\hspace{-0.5mm}-\hspace{-0.5mm}\mathsf{LL}_0}$ reflects the amount of useful information contained in each temporal bin.

Moreover, despite the non-decreasing behavior of $\overline{\mathsf{LL}_F\hspace{-0.5mm}-\hspace{-0.5mm}\mathsf{LL}_0}$, $(a)$ in~\eqref{eq:size_selection} actually decreases for $F>5$ (the left plot of Fig.~\ref{fig:example_3}).
Since a larger $F$ reduces the gap between $P_\mathsf{th}^{(F)}$ and $\psi_F^2$ (Fig.~\ref{fig:example_1}), it contributes to a decrease in the information acquisition probabilities (e.g., $\mathsf{P}_F^{(F)}$ decreases with $F$ in the right plot of Fig.~\ref{fig:example_2}) and results in a reduction in the effective amount of information.
Using the last two rows of Table~\ref{tb:example} (or the left and middle plots of Fig.~\ref{fig:example_3}), we evaluate our cost function values for $F\in[3,8]$ to be $\{0.79,0.87,0.92,0.90,0.88,0.86\}$ (shown in the right plot of Fig.~\ref{fig:example_3}).
As a result, our selection criterion in~\eqref{eq:size_selection} determines $F^\star=5$ to be the number of measurements to be taken for our features, and this is equivalent to the actual number of signal-contained bins $\tilde{F}=5$.
\end{example}

The overall process of our feature size selection can be summarized as follows.
First, for a given positioning scenario, the required information for evaluating the objective function of~\eqref{eq:size_selection} is obtained. 
Then, from a given search range of $F$, the most effective feature size $F^\star$ is determined using~\eqref{eq:size_selection}.
Once $F^\star$ is determined, we train our P-NN using the features consisting of the $F^\star$ largest powers and their temporal locations. 

Note that our feature selection mechanism does not need any prior training of the P-NN. 
Hence, the model training complexity remains the same regardless of the search range of $F$ in~\eqref{eq:size_selection}.
Moreover, our feature size selection is conducted completely offline, which means that our algorithm can be practically adopted into learning-based WP systems without increasing their online operation complexity.
Nevertheless, utilizing the P-NN along with our feature size selection still requires a new set of training data and network training each time there is a considerable change in the localization environment.

\section{Numerical Evaluation}\label{sec:numerical}

We conduct a set of numerical experiments to evaluate the effectiveness of our proposed features and the performance of P-NN.
For the geographical layout, we consider a rectangular sensor space of $d_\mathsf{x}=6$~m, $d_\mathsf{y}=3$~m, and $d_\mathsf{z}=2$~m and and a cylindrical target space of $d_\mathsf{r}=10$~m and $d_\mathsf{h}=4$~m.
We place $M=12$ sensors inside the sensor space to resemble the shape of a vehicle.
Note that we use such a placement of sensors to represent a mobile environment for WP.
For wireless channels, we consider two scenarios from the IEEE UWB standard~\cite{Molisch04}: residential (RES) and outdoor (OUT) environment.
For each scenario, we generate $L$ randomly located channel clusters using Poisson distribution of mean $\overline{L}$ and set $K_l=6$ for all $l$.
Numerical values for the scenario-dependent parameters $\overline{L}$, $\sigma^2_\mathsf{s}$, $\sigma^2_\mathsf{c}$ are given in Table~\ref{tb:scenario_parameter}.
For each channel path, we generate $\mu_{m,l,k}$ using the mean $\overline{\mu}=0.67$~dB and variance $\widetilde{\mu}=0.28$~dB~\cite{Molisch04}.
For the temporal parameters, we set $\kappa=1.5$~ns, $\Gamma=25$~ns, $\gamma=5$~ns.
Regarding the pathloss, we set $\xi=2$ and consider $\overline{P}_m=-45$~dBm and $\overline{d}_m=1$~m for all sensors.
For signal transmission and processing steps, we assume $W=2$ GHz,~$T_\mathsf{f}=200$ ns, and $T_\mathsf{g}=2$ ns to have $N_\mathsf{b}=100$.
For each sensor $m$, we define the SNR as ${\mathbb{E}[\beta_{m,0,0}}]/{\sigma^2_{\mathsf{n},m}}$, where the expectation is over the target space.
To impose the NLOS condition, for each scenario, we remove all existing LOS paths by setting $a_{m,0,k}=0$ for all $m$ and $k$.
For the KL divergence estimation, we use $u=30$.

\begin{figure}[!t]
    \centering
    \begin{subfigure}[!h]{0.49\linewidth}
        \centering
        \includegraphics[width=1\linewidth]{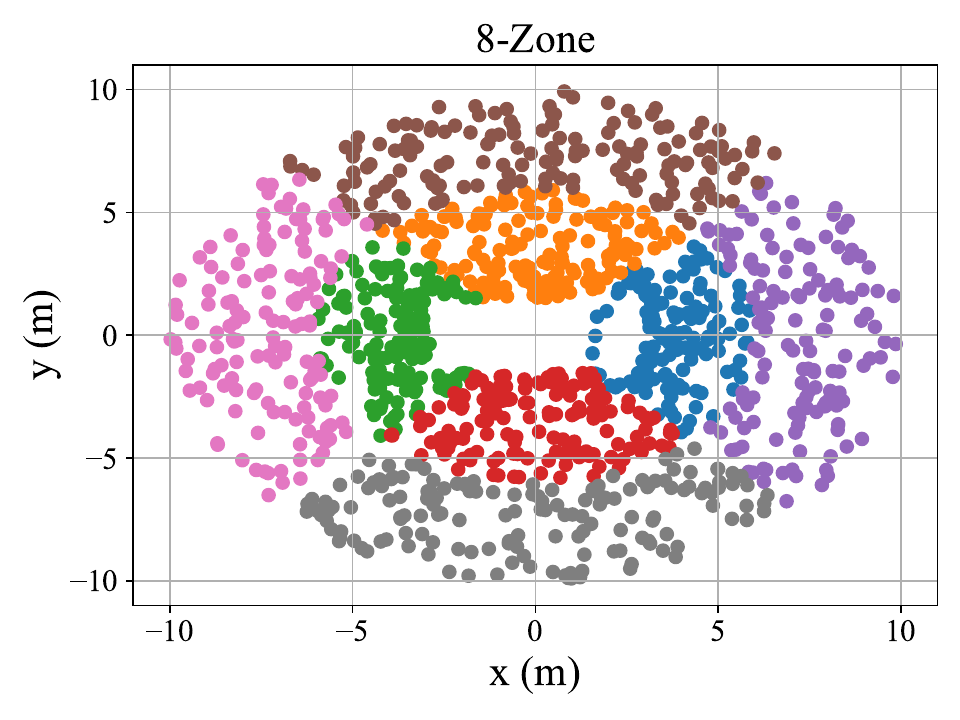}
    \end{subfigure}
    \begin{subfigure}[!h]{0.49\linewidth}
        \centering
        \includegraphics[width=1\linewidth]{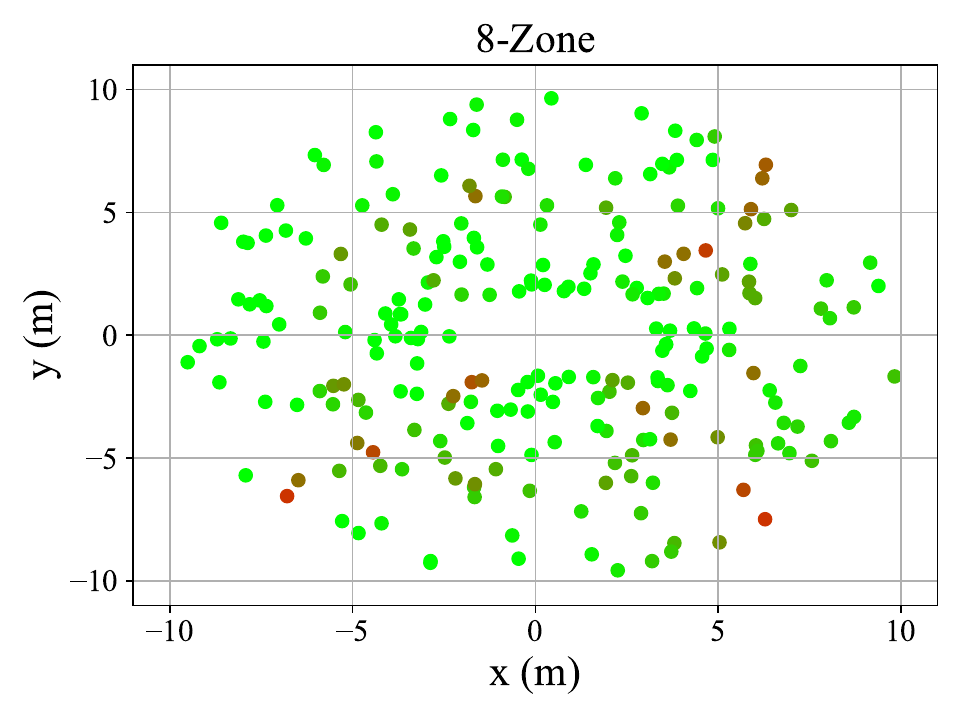}
    \end{subfigure}
    \caption{An illustration of training (left) and testing (right) sets in a 2D plane. For the training set, same color implies the same classification zone. For the testing set, redder color indicates lower classification accuracy.}
    \label{fig:sets}
\end{figure}

For training data, we randomly generate $D=30,000$ target locations inside the target space.
For each target location, the feature $\mathcal{D}^{(i)}$ is generated and paired with a label $\rho_i$.
To train models, we use an Adam optimizer of learning rate $0.001$.
Training is performed over $50$ epochs with the random batch size $256$.
For the testing phase, $6,000$ target locations are randomly generated, and a pair of $\mathcal{D}$ and $\rho$ is obtained for each location.
A visual illustration of our training and testing sets is provided in Fig.~\ref{fig:sets}.
To evaluate the classification performance, we predict $\widehat{\rho}$ for each location in the testing data and compare it with $\rho$.
We consider that a target is correctly positioned only if $\widehat{\rho}=\rho$.
For statistical significance, the result was obtained after averaging over $20$ independent simulation runs and five different scenarios. 

\begin{table}[!t]
\captionsetup{justification=centering, labelsep=newline}
\centering
\setlength\extrarowheight{2pt}
\caption{Simulation parameters for residential (RES) and outdoor (OUT) environments}
\label{tb:scenario_parameter}
\begin{tabular}{|c|c|c|c|} 
 \hline
 Scenario & $\overline{L}$ & $\sigma^2_{\text{s}}$ & $\sigma^2_{\text{c}}$ \tabularnewline
 \hline
  RES & $3$ & $3$ dB & $3$ dB \tabularnewline
 \hline
  OUT & $12$ & $3$ dB & $1$ dB  \tabularnewline
 \hline
\end{tabular}
\vspace{-2mm}
\end{table}

\textbf{Effectiveness of the proposed features:} First, we evaluate the effectiveness of our proposed features.
For comparison, we consider two intuitive baseline approaches to reduce the feature size: (i) taking power measurements from the first $F$ temporal bins (i.e., $n=0,1,\ldots,F-1$) and (ii) taking power measurements from $F$ randomly selected bins.
We use three classic supervised learning algorithms: fully connected layers (FCL) with three $50$-neuron hidden layers, SVM for one-to-rest multi-class classification, and KNN with $k=11$.

\begin{table}[!t]
\centering
\setlength\extrarowheight{3pt}
\setlength\tabcolsep{1.8pt}
\caption{Comparison in 8-zone classification performance by different ways of selecting features. Performance is evaluated by several algorithms: fully connected layers (FCL) with three 50-neuron hidden layers, support vector machine (SVM) for one-to-rest multi-class classification, and $k$-nearest neighbors (KNN) with $k=11$.}
\label{tb:feature}
\begin{tabular}{|c c c|C C C|C C C|C C C|} 
    \hline
    &  &  & \multicolumn{3}{c|}{Proposed} & \multicolumn{3}{c|}{First} & \multicolumn{3}{c|}{Random} \tabularnewline
    \hline
    \multicolumn{2}{|c}{Channel} & $F$ & FCL & SVM & KNN & FCL & SVM & KNN & FCL & SVM & KNN \tabularnewline
    \hhline{|===|===|===|===|}
        &      &  5 & \textbf{89.6} & 88.1 & 63.4 & 44.1 & 42.8 & 42.6 & 35.5 & 33.7 & 29.1 \tabularnewline
        &      & 10 & \textbf{91.1} & 89.6 & 68.9 & 71.8 & 70.3 & 62.6 & 49.2 & 46.4 & 33.1 \tabularnewline
    LOS & 15dB & 15 & \textbf{91.1} & 89.4 & 69.4 & 87.1 & 84.3 & 70.9 & 58.5 & 55.9 & 35.2 \tabularnewline
        &      & 20 & \textbf{90.1} & 89.4 & 69.2 & 88.7 & 86.2 & 66.4 & 65.4 & 61.6 & 36.6 \tabularnewline
        &      & $N_\mathsf{b}$ & \textbf{91.1} & 89.7 & 67.3 & \textbf{91.1} & 89.7 & 67.3 & \textbf{91.1} & 89.7 & 67.3 \tabularnewline
    \hline
        &      &  5 & \textbf{67.9} & 62.3 & 45.7 & 39.3 & 28.9 & 36.9 & 21.1 & 16.6 & 20.2\tabularnewline
        &      & 10 & \textbf{69.4} & 63.4 & 45.4 & 58.5 & 46.7 & 48.0 & 26.9 & 21.2 & 22.7\tabularnewline
    LOS &  5dB & 15 & \textbf{69.6} & 63.7 & 44.2 & 67.7 & 55.7 & 49.5 & 31.2 & 25.6 & 23.9\tabularnewline
        &      & 20 & 70.0 & 64.1 & 43.6 & \textbf{70.5} & 60.8 & 49.3 & 35.5 & 28.6 & 24.7\tabularnewline
        &      & $N_\mathsf{b}$ & \textbf{71.2} & 64.3 & 43.9 & \textbf{71.2} & 64.3 & 43.9 & \textbf{71.2} & 64.3 & 43.9\tabularnewline
    \hline
        &      &  5 & \textbf{79.9} & 73.8 & 50.3 & 12.5 & 12.4 & 12.9 & 29.0 & 23.0 & 27.2\tabularnewline
        &      & 10 & \textbf{81.7} & 76.0 & 53.6 & 14.7 & 14.2 & 15.1 & 38.9 & 31.4 & 30.9\tabularnewline
   NLOS & 15dB & 15 & \textbf{82.1} & 75.9 & 53.7 & 27.3 & 23.7 & 26.5 & 46.0 & 37.4 & 32.1\tabularnewline
        &      & 20 & \textbf{82.2} & 75.5 & 53.6 & 45.7 & 36.5 & 39.7 & 51.3 & 43.6 & 37.4\tabularnewline
        &      & $N_\mathsf{b}$ & \textbf{82.4} & 75.8 & 52.7 & \textbf{82.4} & 75.8 & 52.7 & \textbf{82.4} & 75.8 & 43.6\tabularnewline
    \hline
        &      &  5 & \textbf{47.4} & 41.0 & 29.1 & 12.5 & 12.6 & 12.6 & 16.3 & 14.7 & 16.4\tabularnewline
        &      & 10 & \textbf{48.4} & 41.5 & 27.3 & 13.8 & 13.3 & 13.8 & 19.3 & 16.2 & 17.2\tabularnewline
   NLOS & 5dB  & 15 & \textbf{49.0} & 42.1 & 26.7 & 22.5 & 16.8 & 20.6 & 21.9 & 18.0 & 17.4\tabularnewline
        &      & 20 & \textbf{49.0} & 42.5 & 26.3 & 34.0 & 26.2 & 24.9 & 23.9 & 19.7 & 18.3\tabularnewline
        &      & $N_\mathsf{b}$ & \textbf{50.2} & 42.3 & 27.1 & \textbf{50.2} & 42.3 & 27.1 & \textbf{50.2} & 42.3 & 27.1\tabularnewline
    \hline
\end{tabular}
\end{table}

\begin{figure}[!t]
    \centering
    \includegraphics[width=0.85\linewidth]{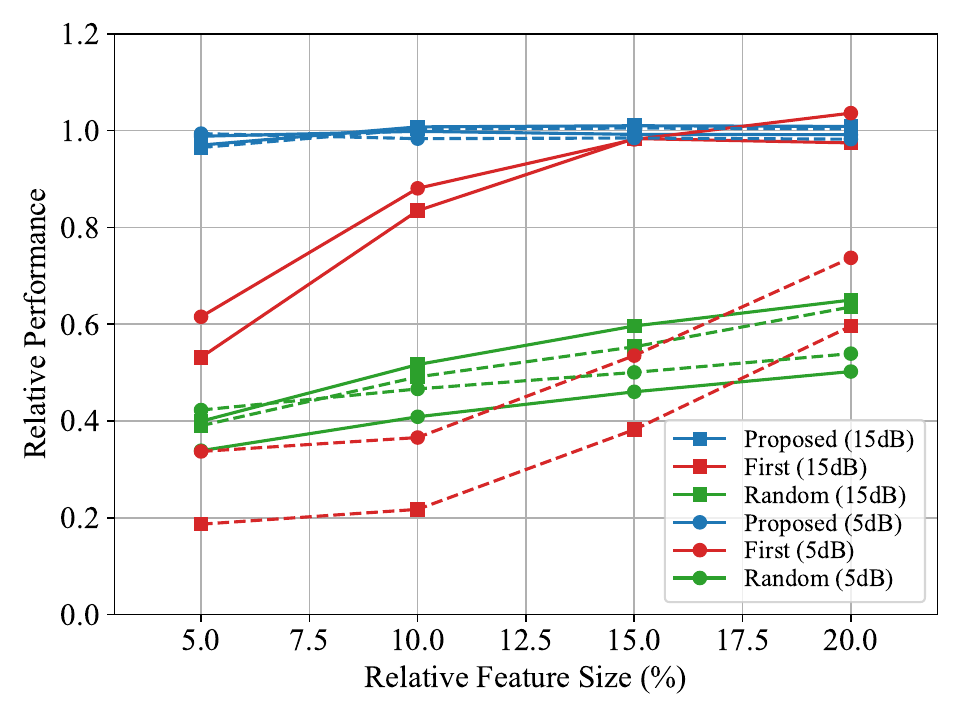}
    \caption{Classification performance vs. feature size plot of different feature size reduction methods. Solid and dashed lines indicate LOS and NLOS conditions, respectively. Performance is normalized to the one obtained using full PDP and averaged over three different classification algorithms: FCL, KNN, and SVM. Feature size is normalized to $N_\mathsf{b}=100$. The proposed features provide robust performance.}
    \label{fig:relative_performance}
    \vspace{-2mm}
\end{figure}

In Table~\ref{tb:feature}, we summarize the classification performance obtained with $N_\mathsf{z}=8$ over a number of different channel conditions.
Note that $F=N_\mathsf{b}$ refers to using full PDP for the features.
From the table, we make the following observations.
First, taking random $F$ measurements yields low performance in general.
This implies that there is a certain set of measurements located across the $N_\mathsf{b}$ temporal bins that are important for WP.
Second, taking the first $F$ bins exhibits a significant performance gap between LOS and NLOS channels.
Since taking the earliest powers is suitable for capturing the LOS path signals, the performance drastically drops for the NLOS channel condition. 
Meanwhile, using our feature set yields both high and robust performance across the channel conditions and algorithms.
Also, for all cases, taking the largest powers can reach the peak performance (i.e., performance with full PDP) within $F=20$.
Particularly, the performance begins to saturate around after $F=10$, with a maximum increase of $0.6\%$ in classification accuracy beyond this point. Therefore, we verify that our proposed feature selection method is able to effectively locate the temporal bins that are significant for WP and reach near-maximum performance with much lower feature size. In other words, our methodology yields improvements in the performance-complexity tradeoff for WP.

In Fig.~\ref{fig:relative_performance}, we provide a classification performance vs. feature size plot for various channel conditions.
To focus on evaluating the performance-efficiency tradeoff, we normalize both the performance and feature size to the case of using full PDP.
From the figure, we see that using the proposed features can achieve performance close to one (i.e., same as using full PDP) even when the feature size is reduced to $10\%$.
Unlike other baselines, which show varying performance depending on the channel conditions, our feature set demonstrates its robustness by keeping the performance high at all conditions.

\begin{table}[!t]
\centering
\setlength\extrarowheight{2pt}
\setlength\tabcolsep{5pt}
\caption{Classification and runtime performance attained from using different power measurement schemes: energy detector (ED) and matched filter (MF). Presented runtime values include only the power measurement steps to obtain $\boldsymbol{\varepsilon}_m$ from $r_m(t)$. Improved performance from MF comes at the cost of having increased runtime.}
\label{tb:power_schemes}
\begin{tabular}{|c|c|c|c|c|c|c|}
    \hline
    \multicolumn{2}{|c|}{} & \multicolumn{4}{c|}{Classification Accuracy} &  \tabularnewline
    \cline{1-6}
    \multicolumn{2}{|c|}{Channel} & \multicolumn{2}{c|}{LOS} & \multicolumn{2}{c|}{NLOS} & Runtime (s)  \tabularnewline
    \cline{1-6}
    \multicolumn{2}{|c|}{SNR} & 15dB & 5dB  & 15dB & 5dB & \tabularnewline
    \hhline{|=|=|=|=|=|=|=|}
    \multirow{3}{*}{ED} & $F=5$ & 89.6 & 67.9 & 79.9 & 47.4 & \multirow{3}{*}{36.8} \tabularnewline
    \cline{2-6}
    & $F=15$ & 91.1 & 69.6 & 82.1 & 49.0 & \tabularnewline
    \cline{2-6}
    & $F=N_\mathsf{b}$ & 91.1 & 71.2 & 82.4 & 50.2 & \tabularnewline
    \hline
    \multirow{3}{*}{MF} & $F=5$ & 90.9 & 84.4 & 85.4 & 67.0 & \multirow{3}{*}{65.7} \tabularnewline
    \cline{2-6}
    & $F=15$ & 92.9 & 85.7 & 87.1 & 68.1 & \tabularnewline
    \cline{2-6}
    & $F=N_\mathsf{b}$ & 92.8 & 86.1 & 87.4 & 69.1 & \tabularnewline
    \hline
\end{tabular}
\end{table}

\textbf{Performance with different power measurement schemes:} Here, we evaluate the classification performance of our proposed features when different power measurements schemes are employed: energy detector (ED) and matched filter (MF).
Unlike ED, MF utilizes a signal template and correlates across the received signal to achieve higher SNRs for the power measurement.
Note that MF requires the Nyquist rate (i.e., the sampling rate of $2W$) and an extra convolution step and therefore yields significantly higher implementation complexities as compared to ED, which operates on a sub-Nyquist rate of $\frac{1}{T_\mathsf{g}}$~\cite{Dardari08}.
With our simulation setting (i.e., $W=2$ GHz and $T_\mathsf{g}=2$ ns), MF requires an eight times faster sampling rate than ED, which may be prohibitive for low-cost sensors.

In Table~\ref{tb:power_schemes}, we show the classification performance obtained over different channel conditions and the values of $F$.
Similar to the result shown in Table~\ref{tb:feature}, for both ED and MF, our features with lower values of $F$ can approach the performance attained when using full PDP.
We observe that the overall performance improves with MF as it relies on the correlation step to increase the SNR after filtering.
Note that more noticeable improvement is shown for both low SNR and NLOS cases, verifying the effectiveness of MF on harsh channel conditions.

Next, to evaluate the performance-complexity tradeoff between ED and MF, we provide the total runtime that takes for each scheme to acquire the PDP vector $\boldsymbol{\varepsilon}_m$ for the entire training data.
We see that MF takes almost double the time ED takes to measure the power of received signals, as MF involves the additional convolution step.
While MF yields better performance than ED, ED shows a clear advantage in both implementation and computational complexities, and therefore, constitutes a desirable power measurement scheme in mobile applications.   

\textbf{Ablation study on P-NN:} Next, we evaluate our P-NN by performing an ablation study on three key components: directing processing (DP) of measurement matrices, spatial processing on a sparse image (SI), and self-attention layer (SA).
In Table~\ref{tb:ablation}, we provide the classification results obtained by five different combinations of the components, where various channel conditions were applied for comprehensive analysis.
From the table, we make several observations.
First, among the three network components we evaluate, SI provides the most improvement (about 10\% increase as compared to DP-only case) in the classification performance.
For all cases, DP+SI+SA alone yields the highest performance, which implies that each component contributes the training/learning ability of P-NN in a cooperative manner.
This is also confirmed by the pattern where a different combination shows a different degree of improvement in the performance.
For instance, DP is shown to be more effective against harsh channel conditions as it brings noticeable performance improvement with low SNR and/or NLOS condition.
On the other hand, SA shows its effectiveness when the channel condition is fairly good (i.e., with high SNR and/or LOS condition).
Hence, our P-NN is effectively trained by our features, and shows improved classification performance by taking different input formats and processing steps.

\begin{table}[!t]
\centering
\setlength\extrarowheight{3pt}
\setlength\tabcolsep{3.6pt}
\caption{Ablation study on the architecture of P-NN in terms of classification performance. Considered components are direct processing (DP) of measurement matrices, sparse image (SI) processing, and self-attention layer (SA). Each component's effectiveness is articulated over different channel conditions.}
\label{tb:ablation}
\begin{tabular}{|c|c|c|c|c|c|c|c|c|c|}
    \hline
    $N_\mathsf{z}$ & \multicolumn{4}{c|}{$8$} & \multicolumn{4}{c|}{$32$}  \tabularnewline
    \hline
    Channel & \multicolumn{2}{c|}{LOS} & \multicolumn{2}{c|}{NLOS} & \multicolumn{2}{c|}{LOS} & \multicolumn{2}{c|}{NLOS}  \tabularnewline
    \hline
    SNR & 15dB & 5dB  & 15dB & 5dB & 15dB & 5dB  & 15dB & 5dB \tabularnewline
    \hhline{|=|=|=|=|=|=|=|=|=|}
    DP & 89.37 & 60.41 & 76.60 &  33.87 & 72.53 & 38.29 & 60.85 & 15.68 \tabularnewline
    \hline
    SI & 93.42 & 69.01 & 86.02 & 41.15 & 83.09 & 47.65 & 70.24 & 20.74 \tabularnewline
    \hline
    DP+SI & 93.61 & 69.52 & \bf{86.98} & \bf{42.13} & 83.89 & \bf{49.66} & \bf{71.93} & \bf{22.40} \tabularnewline
    \hline
    SI+SA & \bf{94.21} & \bf{70.12} & 86.62 & 41.72 & \bf{83.93} & 48.12 & 70.94 & 21.65 \tabularnewline
    \hline
    DP+SI+SA & \bf{94.51} & \bf{70.62} & \bf{87.43} &  \bf{42.66} & \bf{84.33} & \bf{49.85} & \bf{72.62} & \bf{23.17} \tabularnewline
    \hline
\end{tabular}
\end{table}

\textbf{Impact of feature size selection:} Next, we demonstrate the effectiveness of our feature size selection method described in Sec.~\ref{sec:size_selection}.
In Table~\ref{tb:feature_size}, we provide the performance (in zone classification rate) of our P-NN using different values of $F$ over various channel conditions.
We set the search range of $F$ to $[4,10]$ since we gain no significant improvement in performance on further increasing $F$ for this simulation setting, as shown in Table~\ref{tb:feature}. To clarify, other scenarios may produce optimal $F^{\star}$ that are outside of this range; it will vary according to the shape of sensor/target space, the number/location of channel clusters, the SNR, and other conditions that may impact the properties of power delay profile.
For evaluation purposes, here we are training the P-NN and obtaining its test performance for each value of $F$, though as discussed in Sec.~\ref{ssec:selection_criterion}, $F^{\star}$ can be obtained without repeatedly training the network.
We observe that, for all channel conditions, the value of $F$ that approaches the peak performance varies by scenario.
This implies that the desirable feature size for conducting accurate WP is scenario-specific and depends on the condition of channel propagation induced by channel clusters.
For each row, the numerical value in bold indicates the performance obtained using $F^\star$ from our feature size selection method.
We observe that training our P-NN with $F^\star$ can maintain high classification performance with a relatively lower feature size.
In other words, $F^\star$ becomes the point where the marginal increase in classification performance is noticeably reduced.
This verifies that taking the largest power and time measurements constitutes minimum description features for navigating the performance-complexity tradeoff.
Overall, our feature size selection can adaptively determine the dimensions of our features and lead to high WP performance.

\begin{table}[!t]
\centering
\setlength\extrarowheight{1pt}
\setlength\tabcolsep{2.5pt}
\caption{Zone classification rates (in percent) of P-NN with different values of $F$. The rates achieved using $F^\star$ in~\eqref{eq:size_selection} are indicated in bold. We set $\epsilon=0.8 (\text{or }0.6)$ for the LOS (or NLOS) channel scenarios. The value of $F$ that reaches the peak performance varies by scenarios.}
\label{tb:feature_size}
\begin{tabular}{|c|c|c|c|c|c|c|c|c|c|} 
    \hline
    Scenario \# & SNR & $F$ = $4$ & $F$ = $5$ & $F$ = $6$ & $F$ = $7$ & $F$ = $8$ & $F$ = $9$ & $F$=$10$ \tabularnewline
    \hline
    LOS \#3 & \multirow{4}{*}{15dB} & 91.21 & 91.59 & 92.07 & 92.35 & 92.51 & 92.67 & \bf{92.82} \tabularnewline
    \cline{1-1}\cline{3-9}
    LOS \#4 & & 88.21 & 89.42 & 90.11 & \bf{90.51} & 90.88 & 90.84 & 90.89 \tabularnewline
    \cline{1-1}\cline{3-9}
    NLOS \#3 & & 76.31 & 77.25 & 77.79 & \textbf{77.80} & 78.14 & 78.25 & 78.41 \tabularnewline
    \cline{1-1}\cline{3-9}
    NLOS \#4 & & 69.67 & 72.30 & 74.48 & 75.59 & 76.00 & 76.79 & \bf{77.24} \tabularnewline
    \hline
    LOS \#3 & \multirow{4}{*}{5dB} & 68.48 & 69.67 & 70.32 & 70.71 & 71.03 & 71.14 & \bf{71.24} \tabularnewline
    \cline{1-1}\cline{3-9}
    LOS \#4 & & 69.71 & 70.50 & 70.92 & 71.45 & 72.09 & 72.24 & \textbf{72.37} \tabularnewline
    \cline{1-1}\cline{3-9}
    NLOS \#3 & & 44.19 & 44.64 & 44.94 & 45.23 & 45.12 & \bf{45.39} & 45.57 \tabularnewline
    \cline{1-1}\cline{3-9}
    NLOS \#4 & & 49.22 & 49.26 & 49.46 & \textbf{49.80} & 50.00 & 50.21 & 50.15 \tabularnewline
    \hline
\end{tabular}
\end{table}

\begin{figure}[!t]
    \centering
    \includegraphics[width=0.85\linewidth]{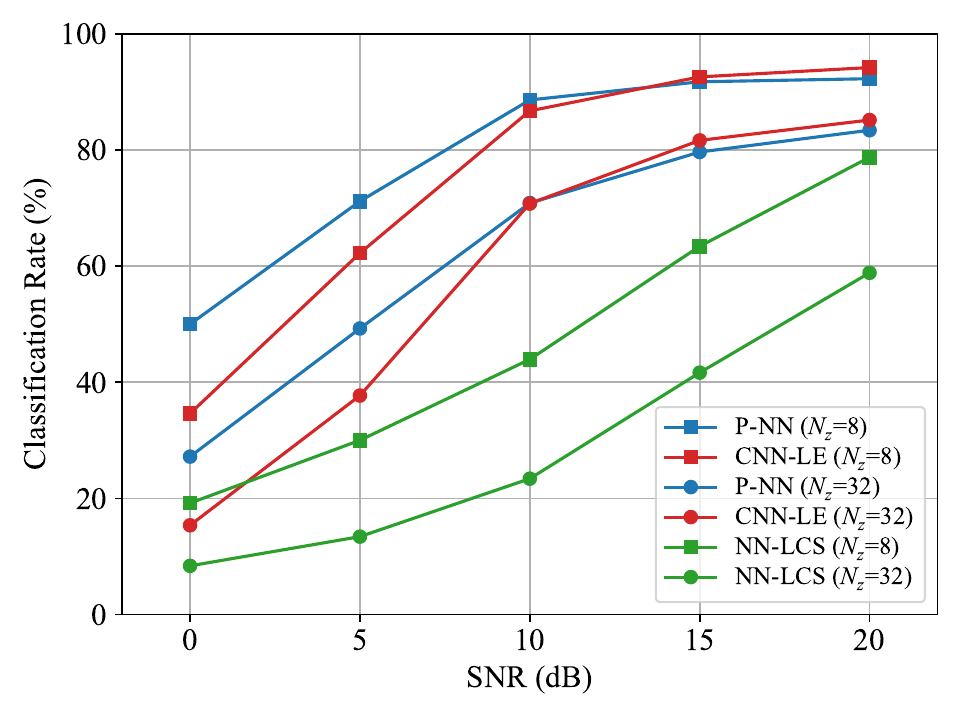}
    \caption{Performance vs. SNR of different WP algorithms with residential LOS channels. Feature sizes for CNN-LE and NN-LCS are $1200$ and $24$, respectively. Feature size for the proposed ranges from $72$ to $240$. The performance advantage of P-NN becomes noticeable in low SNRs.}
    \label{fig:LOS_rate_res}
    \vspace{-1mm}
\end{figure}

\begin{figure}[!t]
    \centering
    \includegraphics[width=0.85\linewidth]{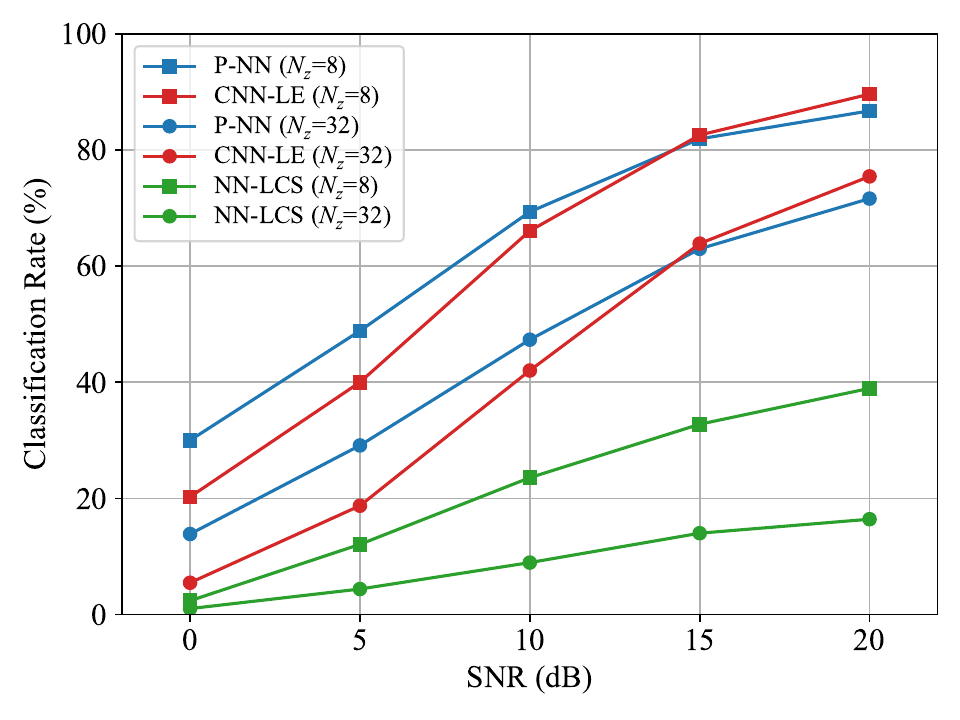}
  \caption{Performance vs. SNR of different WP algorithms with residential NLOS channels. Feature sizes for CNN-LE and NN-LCS are $1200$ and $24$, respectively. Feature size for the proposed ranges from $72$ to $240$. The performance advantage of P-NN becomes noticeable in low SNRs.}
  \label{fig:NLOS_rate_res}
  \vspace{-1mm}
\end{figure}

\textbf{Classification performance of P-NN:} Now we compare the performance of P-NN with the baselines, for which we consider CNN-LE~\cite{Nguyen20} and NN-LCS~\cite{Zheng23}.
CNN-LE is the WP algorithm that takes PDP as input features and utilizes a set of convolutional and maxpooling layers to perform localization.
On the other hand, NN-LCS takes both TOA and RSS measurements and uses FC layers to obtain a set of distance estimation vectors.
Then, the least-squares estimation is applied to estimate the target location.
Compared to CNN-LE, which uses the feature of size $MN_\mathsf{b}$, NN-LCS only takes $2M$ measurements.
We consider CNN-LE and NN-LCS as our baselines since they respectively adopt similar channel model and positioning layout as our work, from which we can provide an objective evaluation and comparison.
For the baselines, we determine the zone classification output based on the coordinates predicted by the algorithms.

\begin{figure}[!t]
    \centering
    \includegraphics[width=0.85\linewidth]{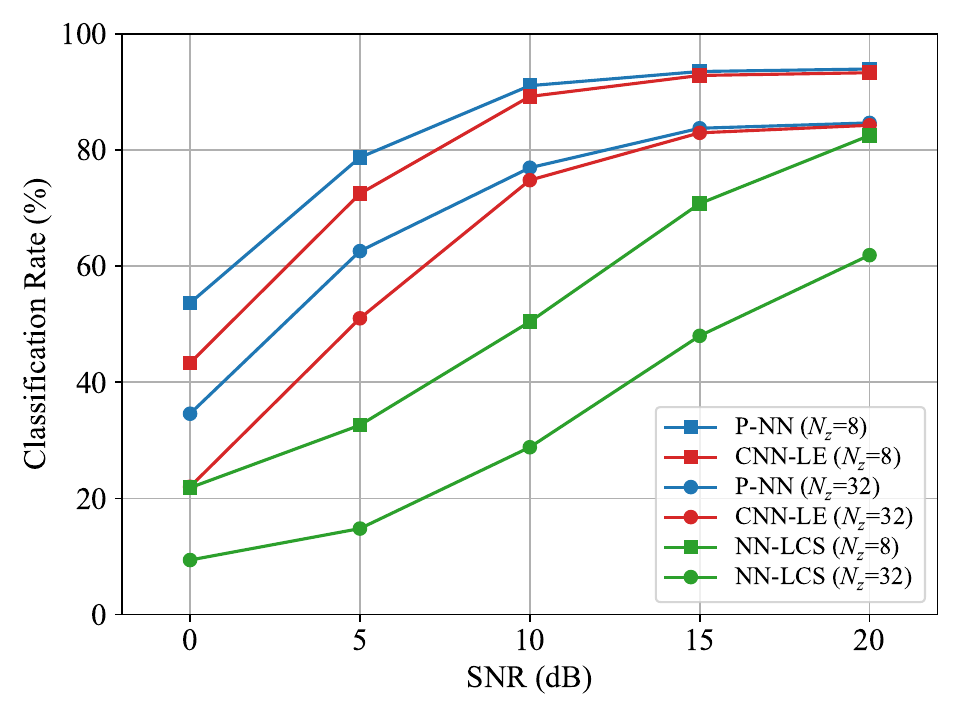}
    \caption{Performance vs. SNR of different WP algorithms with outdoor LOS channels. Feature sizes for CNN-LE and NN-LCS are $1200$ and $24$, respectively. Feature size for the proposed ranges from $72$ to $240$.}
    \label{fig:LOS_rate_out}
    \vspace{-1mm}
\end{figure}

\begin{figure}[!t]
    \centering
    \includegraphics[width=0.85\linewidth]{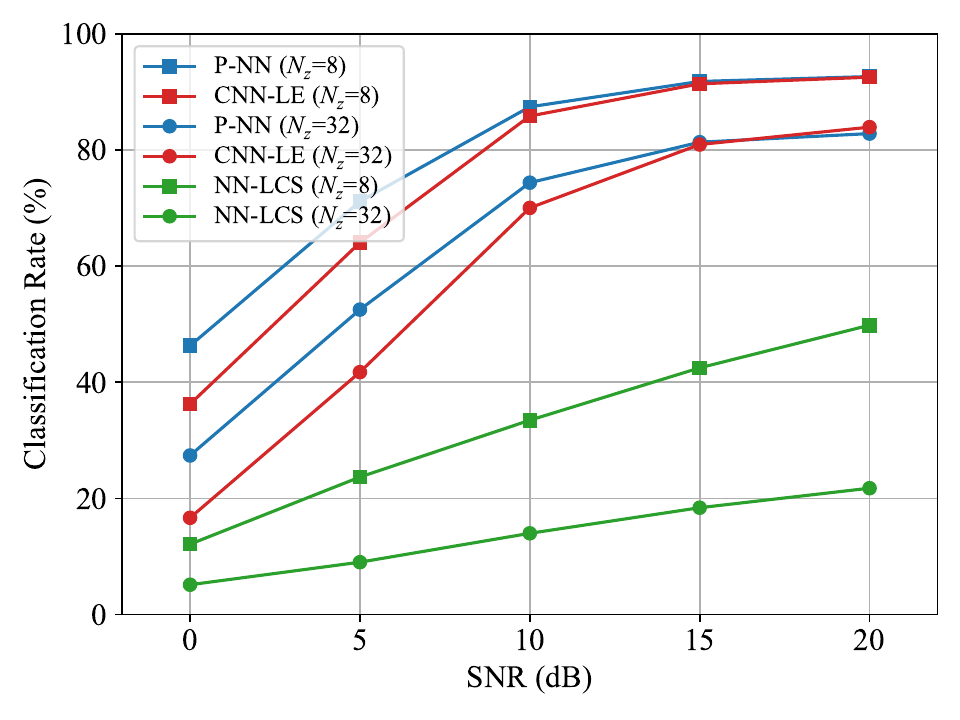}
    \caption{Performance vs. SNR of different WP algorithms with outdoor NLOS channels. Feature sizes for CNN-LE and NN-LCS are $1200$ and $24$, respectively. The feature size for the proposed P-NN ranges from $72$ to $240$.}
    \label{fig:NLOS_rate_out}
    \vspace{-1mm}
\end{figure}

First, we provide classification rate vs. SNR plots for the residential scenario in Figs.~\ref{fig:LOS_rate_res} and~\ref{fig:NLOS_rate_res}. 
For P-NN, we determine $F^\star$ from a range $[4,10]$.
We observe that the performance of NN-LCS in both plots is significantly lower, demonstrating the difficulty of achieving good WP performance from a small-sized feature.
Compared to NN-LCS, both CNN-LE and P-NN provide better performance.
Especially in low SNR, P-NN outperforms CNN-LE as it discards the measurements from noise-only bins, the power of which become greater with low SNR, and thus prevents them from being used in the network training.
In Figs.~\ref{fig:LOS_rate_out} and~\ref{fig:NLOS_rate_out}, we provide performance vs. SNR plots for the outdoor scenario.
We observe that the higher performance is achieved in the outdoor scenario since there are more channel clusters present in the channel space, which provides more channel propagation and signals for the network to utilize.
However, the overall tendency is the same as the residential scenario, where P-NN exhibits the best classification performance.
Given that the performance is competitive between CNN-LE and P-NN (i.e., similar or better performance is achieved depending on the SNR level), our P-NN, which takes only the largest measurements from PDP, takes an advantage in the performance-complexity tradeoff.  

\begin{figure}[!t]
    \centering
    \includegraphics[width=0.85\linewidth]{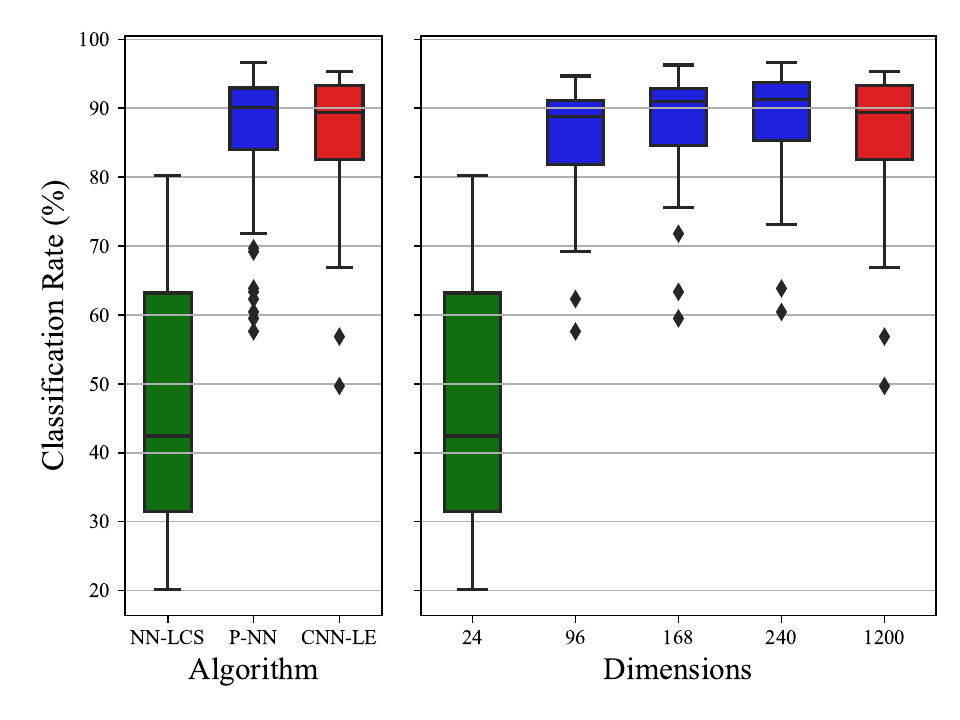}
    \caption{Classification rates obtained with $10$, $15$, and $20$ dB SNRs by different WP algorithms (left) and the number of dimensions (right). For P-NN, we consider $F\in\{4,7,10\}$, leading to the three middle dimensions on the right.}
    \label{fig:tradeoff}
    \vspace{-1mm}
\end{figure}

\textbf{Accuracy range vs. input dimension:} To directly demonstrate the advantage of our P-NN in the performance-complexity tradeoff, we provide box plots showing the range of classification rates obtained by different WP algorithms and the number of feature dimensions in Fig.~\ref{fig:tradeoff}.
We observe that NN-LCS has the lowest dimension, but the performance is low and exhibits a high variance.
CNN-LE exhibits steady and high classification rate, but such a performance is achieved at the cost of utilizing high-dimensional features.
P-NN using our proposed feature set shows a performance similar to the one of CNN-LE at relatively low feature dimensions.
This result demonstrates that our feature set can provide positioning performance that is much more complexity-efficient.

\begin{figure}[!t]
    \centering
    \includegraphics[width=0.85\linewidth]{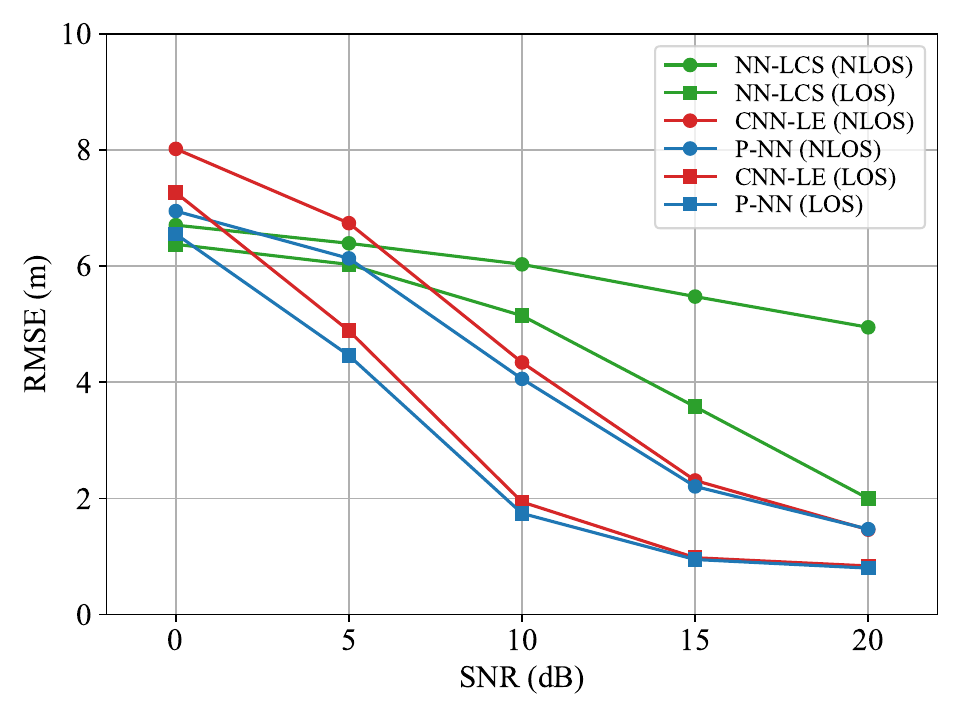}
    \caption{RMSE performance vs. SNR of different WP algorithms with residential channels. Feature sizes for CNN-LE and NN-LCS are $1200$ and $24$, respectively. The feature size for the proposed P-NN ranges from $72$ to $240$.}
    \label{fig:RMSE}
    \vspace{-1mm}
\end{figure}

\textbf{Regression performance of P-NN:} Additionally, we evaluate the regression performance of our P-NN in terms of root mean squared error (RMSE) and compare it with other baselines.
Instead of using the classification layer (i.e., $N_\mathsf{z}$-sized layer with softmax activation), we apply a regression layer that has three neurons with linear activation for estimating 3D coordinates.
If we use $\widehat{\boldsymbol{\ell}}=[\widehat{x},\widehat{y},\widehat{z}]^\top$ to denote the estimated target location of our P-NN, we compute the RMSE performance using the expression $\mathsf{RMSE} = \sqrt{\mathbb{E}\left[(\widehat{x}-x)^2+(\widehat{y}-y)^2+(\widehat{z}-z)^2\right]}$.
In Fig.~\ref{fig:RMSE}, we provide a RMSE versus SNR plot of different WP algorithms evaluated with residential LOS and NLOS channels.
From the figure, we make the following observations.
First, for $10$ dB, $15$ dB, and $20$ dB SNRs, the relative performance across the algorithms is similar to the ones shown in Figs.~$13$ and~$14$, where we evaluate the classification performance.
Hence, our P-NN provides a highly efficient performance-complexity tradeoff for the regression task as well.
Second, for $0$ dB and $5$ dB SNRs, performance of NN-LCS relative to CNN-LE and P-NN is better than what is shown in classification performance.
This implies that, for a regression task, processing the features in an image format is not an effective approach since it is difficult to convey spatial correlation across heavily corrupted measurements from low SNR.
In such a case, providing only the most dominant features in a numerical format (e.g., RSS and TOA values from each sensor) may achieve better performance.
We see that, regardless of SNR levels, our P-NN is able to achieve high performance since its architecture adopts both ways of processing the features.

\section{Conclusion}\label{sec:conclusion}

In this paper, we developed P-NN, a novel technique for WP that utilizes low-dimensional features in mobile settings.
Our minimum description feature set is comprised of a number of largest power measurements and their temporal positions.
For robust performance against various channel conditions, we proposed a method for adapting the feature size by jointly optimizing over the expected amount of information and classification capability, quantified through information-theoretic measures.
Numerical results have shown that using our feature set achieves positioning performance competitive to the one from using PDP, and has a superior performance-complexity tradeoff compared to baseline algorithms.

There are several directions for the future work of this paper.
The feature size selection criterion can be further developed by optimizing the weight parameter to take account the channel conditions.
Also, our feature size selection strategy can be extended to reflect the complexity of network, and thus, can focus on maximizing the learning efficiency.
In addition, more complex WP tasks (e.g., multi-label classification) can be of interest, which may include additional data preprocessing steps to further improve the classification ability of P-NN.
To address the distribution shift problem during online operation, future work can consider integrating drift detection or data monitoring strategies into our WP framework.

\appendices

\section{Proof of Proposition~\ref{prop:greater_F}}\label{appendix:greater_F}

We derive the expression for the change in~\eqref{eq:simple_LL_F} when we increase the number of measurements to take for our features from $F-1$ to $F$ as
\begin{align}
    \widehat{\mathsf{LL}}_F&-\widehat{\mathsf{LL}}_{F-1} = \sum_{n=0}^{F-1}-\frac{\nu}{2}\ln(2\eta^2_{F,n})-\frac{\bar{\varepsilon}^{\text{ord}}_n}{2\eta^2_{F,n}} \nonumber \\ &+\hspace{-1mm}\sum_{n=F}^{N_\mathsf{b}-1}\hspace{-1mm}-\frac{\nu}{2}\ln(2{\psi}^2_F)-\frac{\bar{\varepsilon}^{\text{ord}}_n}{2{\psi}^2_F}+\sum_{n=0}^{F-2}\frac{\nu}{2}\ln(2\eta^2_{F-1,n}) \nonumber \\ &+\frac{\bar{\varepsilon}^{\text{ord}}_n}{2\eta^2_{F-1,n}}+\hspace{-1mm}\sum_{n=F-1}^{N_\mathsf{b}-1}\hspace{-1mm}\frac{\nu}{2}\ln(2{\psi}^2_{F-1})+\frac{\bar{\varepsilon}^{\text{ord}}_n}{2{\psi}^2_{F-1}} \\
    &\hspace{-7mm} = \sum_{n=0}^{F-2}\underbrace{\frac{\nu}{2}\ln \Big(\frac{\eta^2_{F-1,n}}{\eta^2_{F,n}} \Big)+\frac{\bar{\varepsilon}^{\text{ord}}_n}{2} \Big(\frac{1}{\eta^2_{F-1,n}}-\frac{1}{\eta^2_{F,n}} \Big)}_{(a)} \nonumber \\
    &\hspace{-1mm} + \underbrace{\frac{\nu}{2}\ln \Big(\frac{\psi^2_{F-1}}{\eta^2_{F,F-1}} \Big)+\frac{\bar{\varepsilon}^{\text{ord}}_{F-1}}{2} \Big(\frac{1}{\psi^2_{F-1}}-\frac{1}{\eta^2_{F,F-1}} \Big)}_{(b)} \nonumber \\
    &\hspace{-1mm} + \sum_{n=F}^{N_\mathsf{b}-1}\underbrace{\frac{\nu}{2}\ln(\frac{\psi^2_{F-1}}{\psi^2_{F}})+\frac{\bar{\varepsilon}^{\text{ord}}_n}{2} \Big(\frac{1}{\psi^2_{F-1}}-\frac{1}{\psi^2_{F}} \Big)}_{(c)}.
    \label{eq:marginal_LL}
\end{align}
In the last equality, we combine the terms by the index $n$ to separately evaluate the change in the log-likelihood for each temporal bin.

Based on Assumption~\ref{assume:high_SNR} and the definitions made in~\eqref{eq:psi_estimated},~\eqref{eq:lambda_estimated}, and~\eqref{eq:estimated_eta}, for $F\leq\tilde{F}$ we can rewrite~\eqref{eq:psi_estimated} as $\psi^2_F = \frac{1}{\tilde{F}-F}\sum_{n=F}^{\tilde{F}-1}\bar{\varepsilon}^\text{ord}_n$ because $\bar{\varepsilon}^\text{ord}_n$ for $\tilde{F} \leq n \leq N_\mathsf{b}-1$ is negligible.
Since $\bar{\varepsilon}^\text{ord}_{n} \leq \bar{\varepsilon}^\text{ord}_{n-1}$ for $0 \leq n \leq \tilde{F}-1$, we find that 
\begin{equation}
    \psi^2_F \leq \psi^2_{F-1}
    \label{eq:psi_compare}
\end{equation}
always holds when $F\leq \tilde{F}$.
Now, we rewrite the last summation term of~\eqref{eq:marginal_LL} as
\begin{align}
    &\hspace{-3mm}\sum_{n=F}^{N_\mathsf{b}-1}\frac{\nu}{2}\ln \Big(\frac{\psi^2_{F-1}}{\psi^2_{F}} \Big)+\frac{\bar{\varepsilon}^{\text{ord}}_n}{2} \Big(\frac{1}{\psi^2_{F-1}}-\frac{1}{\psi^2_{F}} \Big) \nonumber \\
    &\hspace{-3mm}= \frac{\nu(N_\mathsf{b}-F)}{2}\ln \Big(\frac{\psi^2_{F-1}}{\psi^2_{F}} \Big) + \Big(\frac{\psi^2_{F}-\psi^2_{F-1}}{\psi^2_{F-1}\psi^2_{F}} \Big)\frac{\sum_{n=F}^{\tilde{F}-1}\bar{\varepsilon}^{\text{ord}}_n}{2} \nonumber \\
    &\hspace{-3mm}\leq \frac{\nu(N_\mathsf{b}-F)}{2} \Big(\frac{\psi^2_{F-1}}{\psi^2_{F}}-1 \Big)+ \Big(\frac{\psi^2_{F}-\psi^2_{F-1}}{\psi^2_{F-1}\psi^2_{F}} \Big)\frac{(\tilde{F}-F)\psi^2_{F}}{2} \nonumber \\
    &\hspace{-3mm}=\underbrace{\frac{\nu(N_\mathsf{b}-F)}{2} \Big(\frac{\psi^2_{F-1}\hspace{-1mm}-\psi^2_{F}}{\psi^2_{F}} \Big)}_{(d)}-\underbrace{\frac{(\tilde{F}-F)}{2} \Big(\frac{\psi^2_{F-1}\hspace{-1mm}-\psi^2_{F}}{\psi^2_{F-1}} \Big)}_{(e)},
    \label{eq:marginal_LL_c}
\end{align}
where the inequality holds from $\ln(x)\leq(x-1)$ for $x>0$ and $\psi^2_F = \frac{1}{\tilde{F}-F}\sum_{n=F}^{\tilde{F}-1}\bar{\varepsilon}^\text{ord}_n$.
Since~\eqref{eq:psi_compare} holds, $F\leq\tilde{F}\leq N_\mathsf{b}$, and $\nu=2WT_\mathsf{g}\geq2$ by definition, we observe from~\eqref{eq:marginal_LL_c} that $(d) \geq (e)$, which makes $(c)$ in~\eqref{eq:marginal_LL} always non-negative.
Same derivation steps can be applied for $(a)$ and $(b)$ in~\eqref{eq:marginal_LL}, where we have $\eta^2_{F,n} \leq \eta^2_{F-1,n}$ for any applicable $n$, to show that $(a)$ and $(b)$ are non-negative.
Hence, for $F\leq\tilde{F}$, $\widehat{\mathsf{LL}}_F - \widehat{\mathsf{LL}}_{F-1}$ becomes non-negative, and this makes $\widehat{\mathsf{LL}}_F$ a non-decreasing function of $F$.

Next, based on Assumption~\ref{assume:high_SNR} and the definitions in~\eqref{eq:psi_estimated},~\eqref{eq:lambda_estimated}, and~\eqref{eq:estimated_eta}, for $F>\tilde{F}$ we have (i) $\psi^2_{F}=\psi^2_{F-1}$; (ii) $\lambda_n^{(F)}=\lambda_n^{(F-1)}=\bar{\varepsilon}^{\text{ord}}_n$ for $0\leq n\leq\tilde{F}-1$ and $\lambda_n^{(F)}=\lambda_n^{(F-1)}=0$ for $\tilde{F}-1 < n \leq F$; and (iii) $\eta^2_{F,n}=\eta^2_{F-1,n}=\frac{\bar{\varepsilon}^{\text{ord}}_n}{\sqrt{\nu(2+\nu)}}$ for $0\leq n\leq\tilde{F}-1$ and $\eta^2_{F,n}=\eta^2_{F-1,n}=\psi^2_F$ for $\tilde{F}-1 < n \leq F$. 
If we substitute all these values in~\eqref{eq:marginal_LL}, all $(a)$, $(b)$, and $(c)$ yield to be zero for every value of $n$.
Hence, the difference between $\widehat{\mathsf{LL}}_F$ and $\widehat{\mathsf{LL}}_{F-1}$ becomes zero for $F>\tilde{F}$.
This concludes the proof of the proposition \ref{prop:greater_F}.
\qedsymbol

\bibliographystyle{IEEEtran}
\bibliography{IEEEfull,mybib}

\end{document}